\documentclass[aps,prb,twocolumn,showpacs,superscriptaddress,eqsecnum]{revtex4}
\usepackage{amsbsy,amssymb,amsmath,bm}
\usepackage{graphicx}

\begin{document}

\title{Spin-fermion model with overlapping hot spots and charge modulation
in cuprates.}
\author{Pavel A. Volkov}
\affiliation{Theoretische Physik III, Ruhr-Universit\"{a}t Bochum, D-44780 Bochum, Germany}
\author{Konstantin B.~Efetov}
\affiliation{Theoretische Physik III, Ruhr-Universit\"{a}t Bochum, D-44780 Bochum, Germany}
\affiliation{National University of Science and Technology ``MISiS'', Moscow, 119049,
Russia}
\date{\today }

\begin{abstract}
We study particle-hole instabilities in the framework of the spin-fermion
(SF) model. In contrast to previous studies, we assume that adjacent hot spots
can overlap due to a shallow dispersion of the electron spectrum in the
antinodal region. In addition, we take into account effects of a remnant low
energy and momentum Coulomb interaction. We demonstrate that at sufficiently
small values $|\varepsilon (\pi ,0)-E_{F}|\lesssim \Gamma $, where $E_{F}$
is the Fermi energy, $\varepsilon \left( \pi ,0\right) $ is the energy in
the middle of the Brillouin zone edge, and $\Gamma $ is a characteristic
energy of the fermion-fermion interaction due to the antiferromagnetic
fluctuations, the leading particle-hole instability is a d-form factor Fermi
surface deformation (Pomeranchuk instability) rather than the charge
modulation along the Brillouin zone diagonals predicted within the standard
SF model previously. At lower temperatures, we find that the deformed Fermi
surface is further unstable to formation of a d-form factor charge density
wave (CDW) with a wave vector along the Cu-O-Cu bonds (axes of the Brillouin
zone). We show that the remnant Coulomb interaction enhances the d-form
factor symmetry of the CDW. These findings can explain the robustness of
this order in the cuprates. The approximations made in the paper are
justified by a small parameter that allows one an Eliashberg-like treatment.
Comparison with experiments suggests that in many cuprate compounds the
prerequisites for the proposed scenario are indeed fulfilled and the results
obtained may explain important features of the charge modulations observed
recently.
\end{abstract}

\maketitle

\section{\label{Intr} Introduction}

In the last few years compelling experimental evidence has been gathered for
charge ordering to be a ubiquitous element of the phase diagram of
underdoped cuprates. With different experimental techniques (resonant X-ray
scattering, hard X-ray diffraction, scanning tunneling microscopy(STM)) CDWs
have been directly detected in underdoped samples of YBCO\cite%
{y123REXS-1,y123REXS-2,y123REXS-3,y123REXS-4,y123REXS-5,y123REXS-6,y123XRD-1,y123XRD-2,y123XRD-3,y123XRD-4,y123XRD-5}%
, Bi-2201\cite{bi2201STM-1,bi2201STM-2}, Bi-2212 \cite%
{Bi2212REXS-1,Bi2212REXS-2,Bi2212STM-0,Bi2212STM-1,Bi2212STM-2,Bi2212STM-3}
and, recently, Hg-1201\cite{HgREXS,HgXRD} compounds. Additional input comes
from indirect probes, such as transport measurements \cite%
{y123Transp,HgTransp} and quantum resistance oscillations\cite%
{y123QO-1,y123QO-2,HgQO} consistent with a CDW-like reconstruction \cite%
{seb1,seb2,sachQO} of the Fermi surface (FS), nuclear magnetic resonance\cite%
{y123NMR-1,y123NMR-2}, ultrasound propagation \cite{y123US} and reflectivity
oscillations in pump-probe experiments due to a collective CDW mode \cite%
{y123Refl}.

The charge order revealed by these experiments has several general features.
For a certain doping range the CDW is present in zero magnetic field with
its intensity first appearing below a characteristic temperature $T_{CDW}$
(note that the high-field CDW first observed by NMR\cite{y123NMR-1} in YBCO
has been found\cite{y123NMR-2,y123XRD-5} to be distinct from the zero-field
one). The temperature $T_{CDW}$ exceeds the superconducting transition
temperature $T_{c}$, being generally lower or equal to a pseudogap opening
temperature $T^{\ast }$, such that $T^{\ast }\geq T_{CDW}>T_{c}$. The
intensity increases on cooling down to $T_{c}$, below which it decreases to
a finite value at low temperatures. This picture suggests a competition
between this CDW and superconductivity.

The CDW wave vectors have been universally found\cite%
{y123REXS-5,y123XRD-3,HgXRD,Bi2212STM-0} to be directed along the axes of
the Brillouin zone (axial CDW). The magnitude of the CDW wave vector is
approximately equal for the two orientations and decreases with doping\cite%
{y123XRD-3,bi2201STM-1,Bi2212REXS-2}.

Recent studies have revealed one more feature: the intra-unit-cell structure
of the CDW in Bi-2212\cite{Bi2212STM-2} and YBCO\cite{y123REXS-5} is
characterized by a dominant d-form factor, \textit{i.e.} the charge is
modulated in antiphase at two oxygen sites of the unit cell.

The results of the experiments \cite{Bi2212STM-0,y123REXS-4} also suggest
that the charge ordering is organized in domains where CDW is along one of
the Brillouin zone (BZ) axes only. This contrasts quantum oscillation
experiments\cite{seb1,seb2,sachQO}, where a checkerboard CDW modulated along two
wave vectors simultaneously was used to describe the reconstruction of the
Fermi surface.

All these results clearly distinguish this order from the previously
observed stripe order in La-based cuprates\cite{StripeRev} because the
modulation wave vector for the stripes increases with doping, the
form-factor is predominantly of s' type\cite{y123REXS-6} and the charge
modulation is accompanied by a static spin order not observed in other high-T%
$_{c}$ compounds.

From the theory perspective, various types of CDWs apart from the stripes
were previously considered as a possible explanation for the pseudogap
phenomenon \cite{GapsRev,GapsRev2,EreminLarionov,diCastro1}. Proximity to an incommensurate CDW transition has been also noted to have an effect on the superconducting properties through fluctuations\cite{diCastro2,diCastro3,diCastro4} in a model including electron-phonon interactions, with the signatures of such fluctuations detected in Raman\cite{diCastro5} and ARPES responses\cite{diCastro6}. Recently this topic has reappeared in the context
of the spin-fermion (SF) model, which is known to reasonably reproduce the
d-wave superconducting behavior\cite{SFREV}. In this model a charge order
appears in perturbation theory as a subleading instability\cite{MetSach}
hindered by the curvature of the Fermi surface (however, it has been shown
later in Ref. \onlinecite{SachSau} that the nearest-neighbor Coulomb
interaction favors CDW, offering an explanation for the inequality $%
T_{CDW}>T_{c}$ observed experimentally). This subleading state is formed by
two coexisting d-wave CDW gaps with wave vectors directed along diagonals of
the BZ\cite{Efetov2013,Sach2013} (diagonal CDW). It has been shown\cite%
{Efetov2013} that this order leads to d-form factor charge modulation on
oxygen sites of $CuO_{2}$ plane. It has also been found\cite{Efetov2013}
that the free energy of the charge ordered state can be close enough to the
superconducting (SC) state, such that fluctuations between them destroy both
the orders. At the same time, the gap in the spectrum withstands the
fluctuations and this phenomenon has been used to explain the pseudogap
state in the cuprates.

Several other experimentally relevant predictions have been derived based on
this picture. For example, moderate magnetic fields suppress the
superconductivity and then CDW appears \cite{MEPE} in agreement with the
experiment \cite{y123US}. The core of the vortex should display the charge
modulation \cite{EMPE} and the latter is well seen in, e.g., STM experiments
\cite{hoffman,hamidian}.

It is clear that many features of the charge modulation and its competition
with the superconductivity are well captured in the framework of the SF
model. However, the direction of the CDW modulation vector observed
experimentally in combination with the d-form factor characterizing the
intra-unit-cell charge distribution does not agree with the predictions
derived on the basis of the SF model. Indeed, although the charge modulation
obtained in Refs. \onlinecite{MetSach,Efetov2013,Sach2013} does have the
d-form factor , the modulation vectors obtained there are directed along the
diagonals of BZ, which contrasts the modulation along the BZ axes observed
experimentally.

There have been a number of attempts to approach the problem but the
resolution does not appear to be straightforward. Axial CDW has been deduced
from the SF model in Refs. \onlinecite{cascade,WangChub2014} but the
intra-unit-cell structure obtained in these works possesses a large s-form
factor component. A mixture of the states of Ref. \onlinecite{Efetov2013}
and Ref. \onlinecite{WangChub2014} suggested in Ref. \onlinecite{pepin} does
not correspond to the experiments either because it still contains the
diagonal modulation that is not seen experimentally or the bond modulation
that does not correspond to the d-form factor.

CDW considerations using other models do not explain the robustness of axial
d-form factor CDW in the cuprates either. In Refs. %
\onlinecite{Punk2015,DavisDHLee2013} it has been shown that, provided the
antinodal regions of the Fermi surface are well nested, a
horizontal/vertical instability may become indeed leading but this condition
clearly does not hold in, e.g., Bi-2201, where the Fermi surface
does not show nesting. Mean-field consideration of a three-band model in\cite%
{Kampf2013} leads to the correct direction of the CDW wave vector only for a
closed electron-like Fermi surface, while the charge order with modulation
along the diagonal is dominant for a hole-like FS (which is the case for
underdoped cuprates). The three-band Hubbard model was considered also in
Ref. \onlinecite{yamakawa2015}, where inclusion of vertex corrections led to
the correct direction of the CDW wave vector. However, the obtained form
factor has been found to contain substantial s and s' components. In Ref. %
\onlinecite{Kampf2014} and Refs. \onlinecite{ChowdSach2014,ThomSach} the
pseudogap has been introduced as a separate state related to the parent AF
phase. A qualitative agreement has been obtained for some values of
interaction parameters, while it remains an open question if pseudogap can
be modeled by an AF gap, or FL* state as in Ref. \onlinecite{ChowdSach2014}.

In this paper we extend the treatment of the SF model beyond the vicinities
of eight 'hot spots' to the full antinodal regions of the FS. This is
needed, as is discussed in Section \ref{Constr}, to describe an axial CDW
with a true d-form factor and is also motivated by ARPES data \cite%
{Bi2201ARPES-1,Bi2201ARPES-2,Bi2212ARPES-1,Bi2212ARPES-2} showing that the
energy separation between the hot spots and $(\pi ,0);(0,\pi )$ is actually
quite small. Accordingly, we do not linearize the electron spectrum in the
antinodal regions. In addition to the electron-electron interaction via
paramagnons, we consider also the effects of low-energy part of the Coulomb
interaction, which should not contradict the philosophy of the spin-fermion
model.

Proceeding in this way we show that, provided the antinodal FS is close
enough to the $(\pi ,0)$ and $(0,\pi )$ points, the leading instability in
the d-form factor particle-hole channel is a Fermi surface deformation
(known as Pomeranchuk instability\cite{pomeranchuk,metzner2000,yamase2000}).
The related phase transition occurs at rather high temperatures $%
T_{Pom}\gtrsim T^{\ast }$. We assume that the sample is then reorganized in domains
with broken C$_{4}$ symmetry characterized by different signs of the
Pomeranchuk order parameter. As the hole spectrum remains ungapped at $%
T<T_{Pom}$ the system is susceptible to further instabilities at lower
temperatures. We show that this instability is a CDW with a wave vector
along one of the BZ axes (depending on the sign of the Pomeranchuk order
parameter) in accord with experiments. In the mean field approximation, the
CDW as well as the superconductivity can appear although fluctuations can
mix these states. Thus, the CDW state exists in a form of unidirectional
domains with a correspondingly deformed FS. We point out that, for the
cuprates with experimentally known electron spectra, the FS is, indeed,
sufficiently close to $(\pi ,0);(0,\pi )$, which makes the proposed scenario
applicable to these compounds.

While the Pomeranchuk instability\cite{metzner2000,yamase2000,yamase2005,kee2003} and closely
related electron nematic\cite{kivelson1998,kivelson2011} orders are known
and studied in the context of cuprates (including coexistence with superconductivity\cite{keeSC,yamaseSC} and effects of fluctuations\cite{metznerfluct,yamasefluct}), they have not been considered within
the framework of the SF model and have not been discussed as the reason for
the axial d-form factor CDW robustness.

The Article is organized as follows: in Section \ref{Constr} we discuss the
experimental input justifying the underlying microscopic model, while in
Section \ref{Model} general equations are formulated. In Section \ref{Pom}
we present the leading particle-hole instabilities for a simplified
mean-field model (\ref{Pomsimpl}) yielding analytical results followed by a
consideration of the SF model (\ref{PomSF}) where the solution is obtained
numerically. In Section \ref{CDW} we discuss the emergence of axial CDW for
both the cases. Finally, we discuss in Section\ref{Concl} the obtained
results and their relevance to the charge order in the cuprates.

\section{\label{Constr} Experimental constraints on the microscopic model}

As we are going to use the single band SF model, we should present first of
all a way to relate the quantities we will obtain to observables in the full
$CuO_{2}$ plane. We are mostly interested in the density distribution of
holes at the oxygen sites but these sites are not explicitly present in the
single band SF model. A simple way to relate the density modulation on the $%
O $ atoms to correlation functions of the SF model has been suggested in
Ref. \onlinecite{Efetov2013} (Supplementary Information) where the bond
correlation $\langle c_{i+1}^{\dagger }c_{i}+c_{i}^{\dagger }c_{i+1}\rangle
, $ with $c_{i+1}^{\dagger }$ and $c_{i}$ being electron creation and annihilation
operators on the neighboring $Cu$ sites, was derived to be proportional to
the excess charge density of the $O$ atom located on the bond $(i,i+1).$
This derivation was based on the assumption that holes entered $O$ sites due
to a weak tunneling from $Cu$ sites. Experiments\cite{ZRSexp1} suggest,
though, that doped holes enter mostly $O$ sites, which is not in agreement
with the assumption. Nevertheless, assuming that doped holes form Zhang-Rice
singlets\cite{ZRS} with $Cu$ holes (an assumption that seems to hold well
according to the experiments\cite{ZRSexp2} even in the overdoped regime),
one can come to the same relation between the hole density of the $O$ atoms
and the bond correlation as the one suggested in Ref. \onlinecite{Efetov2013}%
.

In Appendix \ref{app1a} we present a derivation of the formula for the hole
density on $O$ sites in the absence of an on-site modulation on $Cu$ sites
(s-component):
\begin{equation}
\langle p_{j,\sigma }^{\dagger }p_{j,\sigma }\rangle =\frac{p}{4}+\frac{p}{8}%
\langle c_{i+1,\sigma }^{\dagger }c_{i,\sigma }+c_{i,\sigma }^{\dagger
}c_{i+1,\sigma }\rangle _{CO},  \label{ZRSbond}
\end{equation}%
where $p$ is the relative density of doped holes and $p_{j,\sigma }^{\dagger
}$ ($p_{j,\sigma }$) are creation (annihilation) operators for the holes
with spin $\sigma $ on the $O$ atom located between the $Cu$ atoms on sites $%
i$ and $i+1$, and subscript $CO$ means that we write in Eq. (\ref{ZRSbond})
only the contribution due to the charge order. It is useful to have an
analogous expression also in the continuous limit. Considering a $Cu$ atom
at a point $\mathbf{r}$ we write the density at the adjacent $O$ sites in
the $x(y)$ direction as%
\begin{eqnarray}
&&\langle p_{x(y)}^{\sigma \dagger }(\mathbf{r+}\frac{\mathbf{a}_{x(y)}}{2}%
)p_{x(y)}^{\sigma }(\mathbf{r+}\frac{\mathbf{a}_{x(y)}}{2})\rangle =\frac{p}{%
8}n_{0}  \label{a0a} \\
&&+\frac{p}{8}\left\langle c_{\sigma }^{\dagger }\left( \mathbf{r}+%
\mathbf{a}_{x(y)}\right) c_{\sigma }\left( \mathbf{r}\right)
+h.c.\right\rangle _{CO},  \notag
\end{eqnarray}%
where $\mathbf{a}_{x(y)}$ are vectors connecting neighboring $Cu$ atoms ($%
\left\vert \mathbf{a}_{x(y)}\right\vert =a_{0}$)

Finally, we express the electron and hole density modulation on Cu and O sites, respectively, in terms of
the CDW order parameter
\begin{equation}
W_{\mathbf{Q}}(\mathbf{k})=\langle c_{\mathbf{k}-\mathbf{Q}/2,\sigma
}^{\dagger }c_{\mathbf{k}+\mathbf{Q}/2,\sigma }\rangle,  \label{a1}
\end{equation}%
where ${\bf Q}$ is the CDW wave vector and the summation is carried over the BZ, as%
\begin{eqnarray}
\delta n_{Cu}(\mathbf{r}) &=&2e^{i\mathbf{Q}\mathbf{r}}\sum_{\mathbf{k}}W_{\bf Q}(%
\mathbf{k})+c.c,  \label{a2a} \\
\delta n_{O_{x}}(\mathbf{r}) &=&\frac{p}{4}e^{i\mathbf{Q}\mathbf{r}}\sum_{%
\mathbf{k}}\cos (k_{x}a_{0})W_{\bf Q}(\mathbf{k})+c.c.,  \label{a2b} \\
\delta n_{O_{y}}(\mathbf{r}) &=&\frac{p}{4}e^{i\mathbf{Q}\mathbf{r}}\sum_{%
\mathbf{k}}\cos (k_{y}a_{0})W_{\bf Q}(\mathbf{k})+c.c.  \label{a2c}
\end{eqnarray}%
The modulation $\delta n_{Cu}(\mathbf{r})$ corresponds to presence of the
s-form factor component of CDW, the modulation $\delta n_{O_{x}}(\mathbf{r})+\delta
n_{O_{y}}(\mathbf{r})$ gives the s'-form factor component, while $\delta n_{O_{x}}(%
\mathbf{r})-\delta n_{O_{y}}(\mathbf{r})$ stands for the d-form factor component.

Now we can relate the CDW order parameter to the experimental data for the
CDW form factor describing the intra-cell charge distribution. Recent
experiments on BSCCO \cite{Bi2212STM-2} and YBCO \cite{y123REXS-5} systems
demonstrate that the dominant component is the d-form factor one, which implies
that $\delta n_{Cu}=0,\;\delta n_{O_{x}}+\delta n_{O_{y}}=0$. Using Eqs. (%
\ref{a2a}-\ref{a2c}) we write these conditions in the form
\begin{eqnarray}
\sum_{\mathbf{k}}W_{\mathbf{Q}}(\mathbf{k}) &=&0,  \label{d_constr0} \\
\sum_{\mathbf{k}}[\cos (k_{x}a_{0})+\cos (k_{y}a_{0})]W_{\mathbf{Q}}(\mathbf{%
k}) &=&0.  \label{d_constr}
\end{eqnarray}

\begin{figure}[h]
\textbf{\centering
\includegraphics[width=0.5\linewidth]{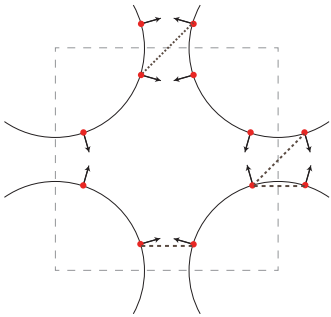} }
\caption{The typical cuprate Fermi surface with eight hot spots. The arrows
represent the direction of the Fermi velocities, while the dashed lines
stand for possible CDW wave vectors (see text).}
\label{fig1a}
\end{figure}

Let us first discuss the constraint (\ref{d_constr0}). The order parameter
of Refs. \onlinecite{MetSach,Efetov2013} with the modulation along the
diagonals of BZ satisfies this condition. In the hot spot approximation, it
follows from the fact that the hot spots connected by the diagonal
wave vector always have antiparallel Fermi velocities (see Fig. \ref{fig1a})
and therefore the magnitudes of the order parameters are equal to each
other. At the same time, this is not true anymore for the CDW wave vector
directed along a BZ axis: the Fermi velocities at the connected hot spots
are no longer antiparallel to each other and moreover, have different angles
for hot spots around $(0,\pi )$ and $(\pi ,0)$. Then, there is no reason for
contributions from hot spots around $(0,\pi )$ and $(\pi ,0)$ to have the
same magnitudes and therefore the constraint (\ref{d_constr0}) is generally
not fulfilled.

In the latter case, the absence of the s-form factor (on-site) component of
the charge modulation encoded in (\ref{d_constr0}) can be understood if one
recalls that there is a strong Coulomb repulsion between the holes on the $%
Cu $ atoms of the $CuO_{2}$ lattice. In effective models, such as the
spin-fermion model, this interaction is assumed to lead to
antiferromagnetism and critical paramagnons \cite{SFREV} when the order is
destroyed. One can come to this result (at least in principle) after
integrating out high energy degrees of freedom. This means, however, that
the low-energy (low-momentum) part of the Coulomb interaction should still
be present in the low-energy effective theory and any additional (quasi)
static modulation of the $Cu$ atoms should therefore cost a considerable
energy. In this situation, a charge distribution without any excess charge
on the $Cu$ atoms can be quite favorable energetically.

\begin{figure}[h]
\textbf{\centering
\includegraphics[width=0.9\linewidth]{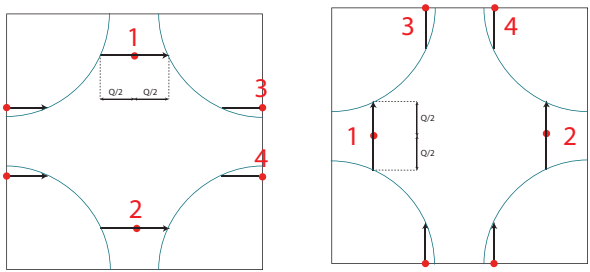} }
\caption{In order for the CDW wave vector ${\bf Q}$ to connect the hot spots, the CDW order parameter W, Eq. (%
\protect\ref{a1}), should be localized at the points 1,2,3 and 4. Left part is for the order parameter of the modulation along horizontal axis, right --- along vertical.}
\label{fig1}
\end{figure}

The s' constraint (\ref{d_constr}) turns out to be even more restrictive for
models that assume that the CDW amplitude is localized in the vicinities of
the points of the Fermi surface connected by the CDW wave vector (hot spots
or, as was recently suggested in \cite{bi2201STM-2}, tips of the Fermi
arcs). It is not difficult to consider the most general case for such models. For both the directions of the CDW there are only 4 such points (Fig. \ref{fig1}), regardless of the two modulation directions coexist (bidirectional) or not (unidirectional).

Taking these 4 relevant points of the FS we attribute the order parameter values $W_{1}$, $W_{2}$, $W_{3}$, and $W_{4}$ to them. Then, the
charge modulations on the atoms with a wave vector ${\bf Q}$ take the simple form (we take ${\bf Q}$ along x for example)

\begin{eqnarray}
\delta n_{Cu} &=&2 e^{i\mathbf{Qr}}(W_{1}+W_{2}+W_{3}+W_{4})+c.c.,
\label{a4} \\
\delta n_{O_{x}} &=&\frac{p}{4} e^{i\mathbf{Qr}%
}(W_{1}+W_{2}-W_{3}-W_{4})+c.c.,  \notag \\
\delta n_{O_{y}} &=-&\frac{p}{4}e^{i\mathbf{Qr}}(W_{1}+W_{2}-W_{3}-W_{4})\cos
(Qa_{0}/2)+c.c.  \notag
\end{eqnarray}

Calculating the modulation amplitudes for different form factors we come to
a rather universal ratio between s' ($\delta n_{s^{\prime }}$) and d ($%
\delta n_{d}$) components:
\begin{equation}
\left| \frac{\delta n_{s'}}{\delta n_{d}} \right|=\frac{1-\cos (Qa_{0}/2)}{1+\cos
(Qa_{0}/2)}  \label{a5}
\end{equation}%
\emph{irrespective of the values of CDW amplitudes at the spots.} This ratio holds for both orientations of the CDW wave vector.

Evaluating the ratio in Eq. (\ref{a5}) for $Q=0.25\ast 2\pi /a_{0}$ (taken
from Ref. \onlinecite{Bi2212STM-2}) we find approximately that it equals $%
0.17$. This result clearly violates the experimental bound $s^{\prime
}/d<1/11.1\approx 0.09$ obtained in Ref.\onlinecite{Bi2212STM-2} indicating
that the order parameter cannot be concentrated in small regions of hot
spots (whatever this term means) and one has to consider contributions
coming from broader regions of BZ. One should note that this experimental
bound is quite conservative because there is no well defined peak observed
in the s' channel.

In contrast, the result for YBCO\cite{y123REXS-5} is different. Taking the
experimental modulation vector $Q=0.31\ast 2\pi /a_{0}$ we obtain the value $%
0.28$ which is approximately equal to the value $0.27$ suggested in Ref. %
\onlinecite{y123REXS-5}. Therefore, this result does not rule out a
possibility that the order parameter in YBCO is localized as a function of
the momentum near certain points of the BZ.

We note that in Ref. \onlinecite{Bi2212STM-2} the function
\begin{equation}
W_{\mathbf{Q}}(\mathbf{k})=B\left[ \cos (k_{x}a_{0})-\cos (k_{y}a_{0})\right]
,  \label{a3}
\end{equation}%
where $B$ is a constant, was used to describe experimental data. This form
of the order parameter clearly obeys both the constraints (\ref{d_constr0}, %
\ref{d_constr}). However, one can show that it contradicts the assumption
that the main contributions come from the vicinity of the Fermi surface. The
absolute value of the order parameter $W_{\mathbf{Q}}\left( \mathbf{k}%
\right) $, Eq. (\ref{a3}), has maxima at the antinodal points $(0,\pi
/a_{0}) $ and $(\pi /a_{0},0)$.
\begin{figure}[h]
\textbf{\centering
\includegraphics[width=0.5\linewidth]{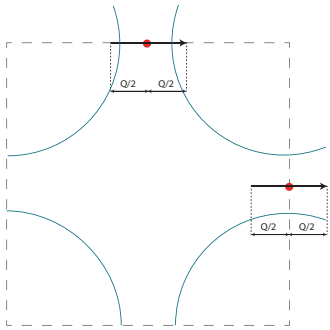} }
\caption{The CDW amplitude from Eq. (\protect\ref{a3}) is maximal at the
dots in the middle of the edges of BZ.}
\label{fig1b}
\end{figure}
Indeed, considering, for example, the x-CDW one is to conclude that, while
the point $(0,\pi /a_{0})$ is located in the middle between the nested parts
of the Fermi surface that can be connected by a vector $\left( Q,0\right) $,
this vector actually connects two points that are located well below the
Fermi surface (see Fig. \ref{fig1b}) in the antinodal region near the point $%
\left( \pi /a_{0},0\right) $ .

The present consideration shows that, in order to obtain a true d-form
factor density wave with a wave vector directed along $x$ or $y$ axes, one
has to formulate the problem in the full antinodal region not restricted to
the vicinities of the points connected by the wave vector. Formulating a
model of interacting fermions one should thus include a possible strong
overlap between different hot spots. This can be done assuming that the band
dispersion near the antinodes is shallow, such that $|\varepsilon (0,\pi
/a_{0})-E_{F}|$ is of the order of the CDW pairing scale ($\varepsilon
\left( k_{x},k_{y}\right) $ stands for the energy spectrum and $E_{F}$ is
the Fermi energy).

Actually, this assumption agrees very well with the results of ARPES
experiments demonstrating that the pseudogap developing in the antinodal
regions can be of the same order of magnitude or larger than $|\varepsilon
(0,\pi /a_{0})-E_{F}|$ (for Bi2201\cite{Bi2201ARPES-1,Bi2201ARPES-2} and
Bi2212\cite{Bi2212ARPES-1,Bi2212ARPES-2}, especially for the antibonding
band). In the framework of the SF model one can associate the pseudogap
scale with the characteristic gap of the model ($b$ in Ref. %
\onlinecite{Efetov2013}). As the gap scale is numerically considerably
smaller than the SF interaction scale $\Gamma$ \cite{Efetov2013}, assuming that $%
|\varepsilon (0,\pi /a_{0})-E_{F}|$ is smaller than $\Gamma$ is a
reasonable assumption for the underdoped cuprates (see also Fig.\ref{figmu}). Moreover, the dependence of the energy scale $\Gamma$ on doping should be weak, unlike the pseudogap and CDW scales which allows to use the approximation above even in the region where pseudogap or CDW become small.

In the next Section we introduce an extended spin-fermion model capable of
taking into account the constraints imposed by the experimental facts.

\section{\label{Model}Extended spin-fermion model and general mean field
equations.}

The original SF model has been written \cite{SFREV} assuming that a Mott
insulator is formed due to a very strong Coulomb repulsion on the $Cu$ atoms
and then destroyed by doping. The philosophy of the SF (semi) phenomenology
is based on integrating out high energy degrees of freedom determining the
antiferromagnetic quantum critical point (QCP). After this procedure is
performed one is left with low-energy fermions and a critical mode
describing antiferromagnetic fluctuations near QCP. There are recent
attempts to derive the SF model from, e.g. t-J model \cite{KochFer}.
Unfortunately, the general effective model derived in that work is still not
sufficiently simple for explicit calculations. Therefore, we prefer to use a
simpler SF model that allows one analytical study in the metallic region
near the QCP.

Using this model one can come to such low energy phenomena as
superconductivity\cite{SFREV}, obtain a CDW instability\cite{MetSach,
Efetov2013}, and study a competition between superconductivity and CDW using
a $\sigma $-model with a composite fluctuating order parameter \cite%
{Efetov2013,MEPE,EMPE}. Calculations based on the $\sigma $-model show that
there is a region of temperatures where only short range correlations of a
mixture of superconducting and CDW orders exist and this region has been
identified with the pseudogap state\cite{Efetov2013}.

All these results have been obtained assuming that important contributions
come only from fermions with momenta close to the Fermi surface. In this
limit one could linearize the spectrum of the fermions, which is a standard
procedure when performing calculations in the weak coupling limit. Moreover,
most important were only small parts of the Fermi surface in the vicinity of
so called \textquotedblleft hot spots\.{\textquotedblright}. It was assumed
also that the Coulomb interaction could only be important for determining
parameters of the low energy effective SF model but it was not present
explicitly there.

Actually, these assumptions are not universally applicable when describing
cuprates and we assume the following new features of the SF model.

\begin{figure}[h]
\textbf{\centering
\includegraphics[trim = 0 150 0 0,scale=0.4]{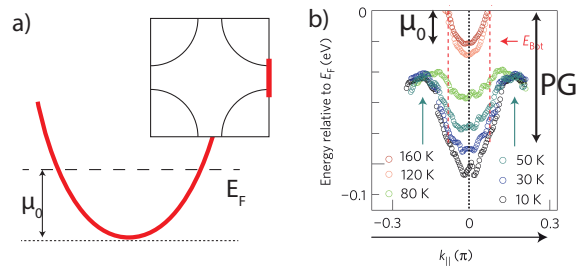} }
\caption{Definition of $\protect\mu _{0}$: a) Cut of the electron dispersion
near the antinode (region shown by the line in the inset) b)Dispersion at
different temperatures obtained from ARPES experiment\protect\cite%
{Bi2201ARPES-1}, with $\protect\mu _{0}$ shown. Note that the pseudogap seen
in the low-temperature dispersions is clearly larger than $\protect\mu _{0}$%
. }
\label{figmu}
\end{figure}

1) Interaction effects are strong not only in the immediate vicinities of
the hot spots but in the full antinodal region. Quantitatively this means
that the energies $\mu _{0}=\left\{ \mathbf{|}\varepsilon (0,\pi
/a_{0})-E_{F}\mathbf{|,\;|}\varepsilon (\pi /a_{0},0)-E_{F}\mathbf{|}%
\right\} $ are smaller or of the same order as the interaction energy scale.
The spectrum $\varepsilon \left( p\right) $ along the edge of the BZ is
represented in Fig. \ref{figmu}.
In reality, the values of $\mu _{0}$ can be smaller than the pseudogap
energy and are of order of several hundred Kelvin \cite%
{Bi2201ARPES-1,Bi2201ARPES-2,Bi2212ARPES-1,Bi2212ARPES-2}. In this
situation, one has to go beyond the vicinity of the FS where the spectrum
cannot be linearized. Indeed, we will see that results concerning the charge
order\cite{MetSach,Efetov2013,Sach2013} drastically change if the spectrum
in the antinodal regions is treated more accurately. Therefore, the
fermionic part of the correct SF model should contain the whole spectrum $%
\varepsilon \left( \mathbf{p}\right) $ of the fermions in the antinodal
regions in order to enable one to consider shallow profiles of the energy
spectra.

2) Following the idea of integrating out high energy degrees of freedom in a
microscopic model there is no reason to neglect at the end the low-energy
and -momentum part of the Coulomb interaction. In contrast, it is quite
natural to treat it together with the critical mode on equal footing. This
can especially be important if a static CDW is formed and the excess charges
interact with each other electrostatically. In addition, the Coulomb
interaction affects the competition between superconductivity and charge
orders reducing the former and enhancing the latter.

The action $S$ for the extended spin fermion model capable to take into
account 1) and 2) can be written in the form
\begin{equation}
S=S_{\mathrm{0}}+S_{\mathrm{\psi }}+S_{\mathrm{\phi }}+S_{\mathrm{c}}
\label{b1}
\end{equation}

In Eq. (\ref{b1}), $S_{0}$ stands for the action of non-interacting fermions
(electrons)%
\begin{equation}
S_{\mathrm{0}}=\int \chi ^{\dagger }\left( X\right) \left[ \partial _{\tau
}+\varepsilon \left( -i\nabla _{\mathbf{r}}\right) -\mu _{0}\right] \chi
\left( X\right) dX,  \label{b2}
\end{equation}%
where $X=\left( \tau ,\mathbf{r}\right) ,$ while
\begin{equation}
S_{\mathrm{\psi }}=\lambda \int \chi ^{\dagger }\left( X\right) \vec{\sigma}%
\vec{\phi}\left( X\right) \chi \left( X\right) dX\mathbf{,}  \label{b2a}
\end{equation}%
describes the interaction of the fermions with the effective exchange field $\vec{\phi}\left( \tau ,\mathbf{r}%
\right) $ of the antiferromagnet. In Eqs. (\ref{b2}, \ref{b2a}), $\chi $ is
the anticommuting fermionic field with two spin components, $\vec{\sigma}$
is the vector of Pauli matrices, and $\tau $ is the imaginary time. The
fermionic field $\chi $ has two spin components
\begin{equation}
\chi =\left(
\begin{array}{c}
\chi _{1} \\
\chi _{2}%
\end{array}%
\right)  \label{b2c}
\end{equation}

The second term in Eq. (\ref{b2}) stands for the fermion energy operator,
and $\mu _{0}$ is the chemical potential counted from $\varepsilon (\pi ,0)$
or $\varepsilon (0,\pi )$. At the moment, we do not make any assumption
about the form of the fermion operator $\varepsilon \left( -i\nabla _{%
\mathbf{r}}\right) $. A proper form of this operator will be chosen when
making explicit calculations.

The Lagrangian $S_{\mathrm{\phi }}$ for the slow exchange field $\vec{\phi}$
is written near QCP as
\begin{equation}
S_{\mathrm{\phi }}=\frac{1}{2}\int \Big[\vec{\phi}\left( X\right) \left[
\hat{D}_{0}^{-1}+\frac{g\vec{\phi}^{2}\left( X\right) }{2}\right] \vec{\phi}%
\left( X\right) \Big]dX.  \label{b3}
\end{equation}%
The propagator $\hat{D}_{0}$ describes the spectrum of the antiferromagnetic
paramagnons near QCP and we chose it in the form
\begin{equation}
\hat{D}_{0}^{-1}=-v_{s}^{-2}\frac{\partial ^{2}}{\partial \tau ^{2}}%
+2a_{0}^{-2}\Big[1-\cos \left[ a_{0}\left( -i\mathbf{\nabla +Q}_{AF}\right) %
\right] \Big]+a.  \label{c6}
\end{equation}%
where $\mathbf{Q}_{AF}=\left( \pi /a_{0},\pi /a_{0}\right) $. In Eq. (\ref%
{c6}), $v_{s}$ is the velocity of the spin waves, $a$ characterizes the
distance from QCP ($a>0$ on the metallic side and $a<0$ in the AF region). We note, that\cite{SFREV} the polarization corrections will strongly affect the paramagnon dynamics at low frequencies, introducing the Landau damping term $\sim |\omega_n|$ into the propagator. In this Section we study the order parameter symmetry properties at the qualitative level and do not consider these. However, they are fully taken into account in the calculations presented in Section \ref{PomSF}.

The last term $S_{\mathrm{c}}$ in Eq. (\ref{b1}) describes the Coulomb
interaction between slowly fluctuating charges. We write this term in the
form%
\begin{eqnarray}
&&S_{\mathrm{c}}=\frac{1}{2}\int V_{\mathrm{c}}\left( X-X^{\prime }\right)
\label{b2b} \\
&&\times \left( \chi ^{\dagger }\left( X\right) \chi \left( X\right) \right)
\left( \chi ^{\dagger }\left( X^{\prime }\right) \chi \left( X^{\prime
}\right) \right) dXdX^{\prime }.  \notag
\end{eqnarray}
As the high energies and momenta are assumed to have been integrated out,
the interaction $V_{\mathrm{c}}\left( X-X^{\prime }\right) $ in Eq. (\ref%
{b2b}) is a low-energy part of the screened Coulomb interaction. It slowly
varies on atomic scales as a function of coordinates and times and vanishes
for fast variations.

Neglecting the quartic interaction and averaging over $\vec{\phi}\left( \tau
,\mathbf{r}\right) $ with the help of Eq. (\ref{b3}, \ref{c6}) one comes to
the effective fermion-fermion interaction due to exchange by the paramagnons%
\begin{eqnarray}
&&S_{\mathrm{int}}=-\frac{\lambda ^{2}}{2}\int D_{0}^{-1}\left( X-X^{\prime
}\right)  \label{b4} \\
&&\times \left( \chi ^{\dagger }\left( X\right) \vec{\sigma}\chi \left(
X\right) \right) \left( \chi ^{\dagger }\left( X^{\prime }\right) \vec{\sigma%
}\chi \left( X^{\prime }\right) \right) dXdX^{\prime }  \notag
\end{eqnarray}%
where $X=\left( \tau ,\mathbf{r}\right) $.

The total fermion-fermion interaction $S_{\mathrm{tot}}$ equals%
\begin{equation}
S_{\mathrm{tot}}=S_{\mathrm{int}}+S_{\mathrm{c}},  \label{b5}
\end{equation}%
and the partition function $Z$ for the model introduced takes the form
\begin{equation}
Z=\int \exp \left[ -S_{\mathrm{0}}\left[ \chi \right] -S_{\mathrm{tot}}\left[
\chi \right] \right] D\chi .  \label{b6}
\end{equation}

Considering both charge orders and superconductivity on equal footing is
convenient with the help of vectors $\Psi $ defined as\cite{Efetov2013}

\begin{equation}
\Psi =\frac{1}{\sqrt{2}}\left(
\begin{array}{c}
\chi ^{\ast } \\
i\sigma _{2}\chi%
\end{array}%
\right) ,\quad \Psi ^{\dagger }=\frac{1}{\sqrt{2}}\left(
\begin{array}{cc}
-\chi ^{t} & -\chi ^{\dagger }i\sigma _{2}%
\end{array}%
\right) ,  \label{ap4b}
\end{equation}
where \textquotedblleft $t$\textquotedblright\ stands for transposition.
This is a standard Gor'kov-Nambu representation and $\chi $ is the
two-component spinor, Eq. (\ref{b2c}).

Then, one can come to an order parameter
\begin{equation}
Q\left( X,X^{\prime }\right) =\left\langle \Psi \left( X\right) \Psi
^{\dagger }\left( X^{\prime }\right) \tau _{3}\right\rangle ,  \label{ap4c}
\end{equation}%
($\tau _{3}$ is Pauli matrix in the Gor'kov-Nambu space) containing pairings
in both particle-hole and superconducting channels. Singlet pairing is most
energetically favorable and in this case one can represent the order
parameter in a form of a $4\times 4$ matrix
\begin{equation}
\mathcal{M}=I_{\sigma }\otimes \hat{M}  \label{ap4da}
\end{equation}%
\begin{equation}
\hat{M}\left( X,X^{\prime }\right) =\tau _{3}\left( \hat{W}\left(
X,X^{\prime }\right) +\hat{\Delta}\left( X,X^{\prime }\right) \right) .
\label{ap4d}
\end{equation}%
In Eq. (\ref{ap4d}) matrices $\hat{W}\left( X,X^{\prime }\right) $ and $\hat{%
\Delta}\left( X,X^{\prime }\right) $ equal,%
\begin{equation}
\hat{W}\left( X,X^{\prime }\right) =\left(
\begin{array}{cc}
W^{\ast }\left( X,X^{\prime }\right) & 0 \\
0 & W\left( X,X^{\prime }\right)%
\end{array}%
\right)  \label{k5}
\end{equation}%
and
\begin{equation}
\hat{\Delta}\left( X,X^{\prime }\right) =\left(
\begin{array}{cc}
0 & \Delta ^{\ast }\left( X,X^{\prime }\right) \\
-\Delta \left( X,X^{\prime }\right) & 0%
\end{array}%
\right)  \label{k6}
\end{equation}%
where $\Delta \left( X,X^{\prime }\right) =\Delta \left( X^{\prime
},X\right) $ and $W\left( X,X^{\prime }\right) =W^{\ast }\left( X^{\prime
},X\right) $ are order parameters for singlet supercoductivity and charge
modulation, respectively, and $I_{\sigma }$ is the unit matrix in the spin
space. We will see later that both the order parameters have d-wave symmetry.

A detailed description of the charge and superconducting orders can be
performed decoupling the electron-paramagnon and Coulomb interactions with a
Hubbard-Stratonovich transformation. As a result of this decoupling, one can
reduce the original integral $Z$ over the fermionic fields, Eq. (\ref{b6}),
to an integral over slowly varying in space and time matrices $Q$.
Calculation of the latter integrals is carried out by finding saddle points
determined by mean field equations and integrating over fluctuations near
these points.

Details of this calculation are presented in Appendix \ref{app0}. Here we
write only the mean field equations keeping the matrix form of the order
parameter $Q\left( X,X^{\prime }\right) $
\begin{eqnarray}
&&-\hat{M}\left( X,X^{\prime }\right) =\delta \left( X-X^{\prime }\right)
\tau _{3}\int V_{\mathrm{c}}\left( X-X_{1}\right)  \notag \\
&&\times \mathrm{tr}\left[ \tau _{3}G\left( X_{1},X_{1}\right) \right]
dX_{1}-V_{\mathrm{c}}\left( X-X^{\prime }\right) \tau _{3}G\left(
X,X^{\prime }\right) \tau _{3}  \notag \\
&&+3\lambda ^{2}D\left( X-X^{\prime }\right) G\left( X,X^{\prime }\right) ,
\label{k1}
\end{eqnarray}%
where the Green's function $G\left( X,X^{\prime }\right) $ satisfies the
equation%
\begin{equation}
\left( \hat{H}_{0}-\hat{M}\right) G\left( X,X^{\prime }\right) =-\delta
\left( X-X^{\prime }\right) ,  \label{k2}
\end{equation}
\begin{equation}
\hat{H}_{0}=\partial _{\tau }-\left[ \varepsilon \left( -i\tau _{3}\nabla _{%
\mathbf{r}}\right) -\mu_0 \right] \tau _{3},  \label{k3}
\end{equation}%
and $D$ is a propagator of critical excitations screened by electron-hole
fluctuations.

The symbol $\mathrm{tr}$ stands for the trace over elements of the matrix $M$%
. In the absence of the Coulomb interaction $V_{\mathrm{c}}\left(
X-X^{\prime }\right) $, Eqs. (\ref{k1}-\ref{k3}) correspond to those derived
in Ref. \onlinecite{Efetov2013}. If, in addition, one linearizes the
spectrum near the Fermi surface the matrix $\tau _{3}$ completely drops out
from Eq. (\ref{k1}, \ref{k2}) and the system becomes degenerate with respect
to superconducting and charge modulation states (matrix elements $\Delta $
and $W$). Fluctuations between these states can be strong leading to the
pseudogap state \cite{Efetov2013}.

However, in the mean field approximation, the charge modulation states and
superconductivity are generally not degenerate and can be considered
separately. Taking the off-diagonal part of the matrix $\hat{M}\left(
X,X^{\prime }\right) $ we obtain for the superconducting order parameter%
\begin{eqnarray}
&&\hat{\Delta}\left( X,X^{\prime }\right) =\frac{1}{2}\left[ 3\lambda
^{2}D\left( X-X^{\prime }\right) +V_{\mathrm{c}}\left( X-X^{\prime }\right) %
\right]  \notag \\
&&\times \left[ \left( \hat{H}_{0}\tau _{3}+\hat{\Delta}\right) ^{-1}-\left(
\hat{H}_{0}\tau _{3}-\hat{\Delta}\right) ^{-1}\right] _{X,X^{\prime }}.
\label{k7}
\end{eqnarray}

Eq. (\ref{k7}) has solutions for d-wave symmetry of the order parameter. The
low-energy part of the Coulomb interaction $V_{\mathrm{c}}\left( X-X^{\prime
}\right) $ in Eq. (\ref{k7}) hinders the superconductivity. At the same
time, the non-linearity in the spectrum obstructs the charge modulation and
one can expect a competition between these two states.

As concerns the charge modulation, we write the equation for $\hat{W}\left(
X,X^{\prime }\right) $ in the form
\begin{eqnarray}
&&\hat{W}\left( X,X^{\prime }\right) =\delta \left( X-X^{\prime }\right)
\int V_{\mathrm{c}}\left( X-X_{1}\right)  \label{k8} \\
&&\times \mathrm{tr}\left[ \left( H_{0}\tau _{3}-\hat{W}\right) ^{-1}\right]
_{X_{1},X_{1}}dX_{1}  \notag \\
&&+\left[ 3\lambda ^{2}D\left( X-X^{\prime }\right) -V_{\mathrm{c}}\left(
X-X^{\prime }\right) \right] \left[ \left( H_{0}\tau _{3}-\hat{W}\right)
^{-1}\right] _{X,X^{\prime }}  \notag
\end{eqnarray}

The quadrupole density wave with the diagonal modulation \cite%
{MetSach,Efetov2013} has already the d-wave symmetry and the first term in
R.H.S. of Eq. (\ref{k8}) (Hartree-type of the contribution) vanishes in this
case. At the same time, states with a charge modulations along the bonds are
very sensitive to the classical part of the Coulomb interaction described by
this term. The expression under the trace is the full excess charge in the
unit cell and it is energetically favorable to have quadrupole-like
configurations for which this term vanishes. It is the first term in R.H.S.
of Eq. (\ref{k8}) that generally leads to the d-form factor symmetry of the charge
modulation even if the latter is not directed along the diagonals of the
lattice. The sign of the Fock-type of the contributions of the Coulomb
interaction $V_{\mathrm{c}}\left( X-X^{\prime }\right) $ in Eq. (\ref{k8})
is opposite to the one in Eq. (\ref{k7}) and this interaction enhances the
charge modulation.

Eqs. (\ref{k7}, \ref{k8}) can be rewritten in the momentum and frequency
representation and solved explicitly. Depending on the parameters, Eqs. (\ref%
{k7}, \ref{k8}) allow not only solutions for superconductivity and charge
modulation with a finite vector $\mathbf{Q}$ but also an intracell charge
modulation with $\mathbf{Q}=0$ leading to a reconstruction of the FS. The
latter corresponds to a Pomeranchuk instability and we will demonstrate in
the next Section that it can be in a certain region of the parameters the
strongest instability in the system. We also note that in this Section we have not considered the normal-state interaction effects, e.g. fermion self-energy and polarization effects. These effects turn out to be important and they are fully taken into account in Section \ref{PomSF}.

\section{\label{Pom}Pomeranchuk instability and reconstruction of the Fermi
surface.}

We concentrate now on studying the charge order formation in the
\textquotedblleft hot regions\textquotedblright\ of the FS approximately
connected by the antiferromagnetic wave vector $(\pi /a_{0},\pi /a_{0})$
(see Fig. \ref{fig2}). At the moment, we neglect the possibility of the
superconducting transition. This assumption can be justified by the presence
of the long range part of the Coulomb interaction in Eqs. (\ref{k7}, \ref{k8}%
). We assume that it is essential only inside the hot regions, thus
enhancing the charge modulation, Eq. (\ref{k8}), and hindering the
superconductivity, Eq. (\ref{k7}).

It is important to emphasize that we consider the situation when eight hot
spots of the traditional SF model \cite{SFREV} strongly overlap due to the
shallow profile of the spectrum near the antinodes and we have effectively
two hot regions 1 and 2 (see Fig. \ref{fig2}). In order to simplify the
calculations, we assume that the Coulomb interaction is large at small
momenta and the Hartree contribution in Eq. (\ref{k8}) is very large unless
the excess intra-unit-cell charge in the language of the one band model is
equal to zero. The latter condition excludes any charge on the $Cu$ atoms
or, in other words, the s-form factor component of the charge distribution.
Moreover, the s'-component can also be neglected due to smallness of ($\cos
k_{x}+\cos k_{y})$ in the antinodal region.

Therefore, we can analyze the mean field equation (\ref{k8}) assuming from
the beginning the d-form factor symmetry of the charge modulations. Of course, the
solution of Ref. \cite{Efetov2013} automatically satisfies this condition.
However, we will see in this section that it is not always most
energetically favorable. We proceed with seeking for solutions of Eq. (\ref%
{k8}) for arbitrary modulation vectors $\mathbf{Q}$ and find the one
providing the highest transition temperature. As the interaction of
electrons via paramagnons is frequency dependent, it is not possible to
obtain a full analytical solution. Moreover, there one needs to modify the
equation (\ref{k8}) to include the normal-state self energy corrections.
This will be done essentially as in Ref. \cite{Efetov2013}. However, the
mechanism of the charge order formation can be understood analytically
considering a simplified model with a frequency and momentum independent
inter-region repulsion replacing the original electron-electron interaction
via paramagnons. This study is presented in the next Subsection \ref%
{Pomsimpl} followed by a numerical investigation of the original SF model in
Subsection \ref{PomSF}.

\subsection{\label{Pomsimpl}Simplified Model}

\begin{figure}[h]
\centering
\includegraphics[width=0.4\linewidth]{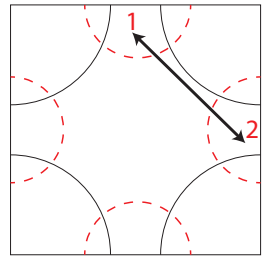}
\caption{BZ with regions 1 and 2.}
\label{fig2}
\end{figure}

We consider two regions of the Fermi surface surrounding the antinodes 1 and
2 in Fig. \ref{fig2}.
We assume a momentum and frequency independent repulsive interregion
interaction and consider only d-form factor charge instabilities with an
arbitrary wave vector $\mathbf{Q}$
\begin{equation}
W_{1}=\langle \chi _{\mathbf{p}+\mathbf{Q}/2}^{1\ast }\chi _{\mathbf{p}-%
\mathbf{Q}/2}^{1}\rangle =-W_{2}=-\langle \chi _{\mathbf{p}+\mathbf{Q}%
/2}^{2\ast }\chi _{\mathbf{p}-\mathbf{Q}/2}^{2}\rangle .  \label{k9}
\end{equation}

The form of the order parameter specified by Eq. (\ref{k9}) guarantees
d-form factor and thus evades the on-site repulsion (first term in R.H.S. of
Eq. (\ref{k8})). We can solve the mean field equation (\ref{k8}) for $W_{%
\mathbf{Q}}$ for an arbitrary modulation vector $Q$.

As we are interested in the situation when the Fermi energy is not far from
the $(0,\pi ),(\pi ,0)$ points in the electron spectrum, we expand the
energy $\varepsilon ^{1,2}(\mathbf{p})$ of the original Hamiltonian, Eq. (%
\ref{k3}) around them%
\begin{equation}
\varepsilon _{p}^{1}=\alpha p_{x}^{2}-\beta p_{y}^{2}-\mu _{0},\;\varepsilon
_{p}^{2}=\alpha p_{y}^{2}-\beta p_{x}^{2}-\mu _{0}  \label{k10}
\end{equation}%
where $\mu _{0}$ is the Fermi energy counted from the saddle points.
Moreover, we will use an averaged dispersion over $p_{y(x)}$ for region
1(2). This is justified if $\beta \ll \alpha $ (small curvature) but one can
also study the qualitative effect of increasing the curvature within this
approximation. Then, we write the effective dispersion in the form
\begin{equation}
\varepsilon _{p}^{1}=\alpha p_{x}^{2}-\mu ,\quad \varepsilon _{p}^{2}=\alpha
p_{y}^{2}-\mu ,  \label{tm_disp}
\end{equation}%
where $\mu =\mu _{0}+\langle \beta p_{\parallel }^{2}\rangle $. We use Eq. (%
\ref{tm_disp}) instead of Eq. (\ref{k10}) because this approximation
simplifies the calculations but, at the same time, we do not expect an
essential difference of results even when $\beta $ is of the same order as $%
\alpha .$ We see that the presence of the curvature leads to an effective
increase of the energy $\mu $ in the antinodal regions. Having in mind
applications to cuprates one can say that $\mu $ varies with the hole doping
decreasing down to the point where the FS closes becoming electron-like ($%
\mu _{0}=0$) and further as one goes into the overdoped regime. The
parameter $\alpha $ can, on the contrary, be considered as independent of
the doping. The mean field Lagrangian $L_{\text{\textrm{M}}\mathrm{F}}$ for
the described model can be written in the form 
\begin{eqnarray}
&&L_{\text{\textrm{M}}\mathrm{F}}=\frac{|W_{\mathbf{Q}}|^{2}}{\lambda _{0}}%
+\sum_{\mathbf{p},\nu =1,2}\Big[\chi _{\mathbf{p}}^{\nu \dagger }\left[
\partial _{\tau }+\varepsilon _{i}\left( \mathbf{p}\right) \right] \chi _{%
\mathbf{p}}^{\nu }  \label{tm_h} \\
&&+\left( -1\right) ^{\nu -1}\left( W_{\mathbf{Q}}^{\ast }\chi _{\mathbf{p}+%
\mathbf{Q}/2}^{\nu \dagger }\chi _{\mathbf{p}-\mathbf{Q}/2}^{\nu }+W_{%
\mathbf{Q}}\chi _{\mathbf{p}-\mathbf{Q}/2}^{\nu \dagger }\chi _{\mathbf{p}+%
\mathbf{Q}/2}^{\nu }\right) \Big], \notag
\end{eqnarray}
where $W_{{\bf Q}}\equiv W_2$.

The charge instabilities can rather easily be analyzed in this model
minimizing the free energy with respect to the order parameter and
linearizing the obtained equations near the transition point $T_{CO}$. Then,
we come to a simple equation replacing Eq. (\ref{k8}) for the model
considered here
\begin{equation}
\frac{1}{2}\sum_{\mathbf{p,}\nu =1,2}\left\{ \frac{\tanh {\frac{\varepsilon
_{\mathbf{p}+\mathbf{Q}/2}^{\nu }}{2T_{CO}}}-\tanh {\frac{\varepsilon _{%
\mathbf{p}-\mathbf{Q}/2}^{\nu }}{2T_{CO}}}}{\varepsilon _{\mathbf{p}+\mathbf{%
Q}/2}^{\nu }-\varepsilon _{\mathbf{p}-\mathbf{Q}/2}^{\nu }}\right\} =\frac{2%
}{\lambda _{0}}  \label{tm_tc}
\end{equation}

Our goal is to find the modulation vector $\mathbf{Q}=\left(
Q_{x},Q_{y}\right) $ yielding the maximal $T_{CO}$ for different values $%
\lambda _{0},\alpha ,\mu $. First, one can notice using Eq. (\ref{tm_disp})
that the L.H.S. of (\ref{tm_tc}) is a sum of two identical functions with
different arguments $f(Q_{x},T_{CO})+f(Q_{y},T_{CO})$, where the first term
corresponds to $\nu =1$ and the second one to $\nu =2$. The temperature $%
T_{CO}^{max}$ is finite because $f$ is a decreasing function of $T_{CO}$ for
any $Q$ for sufficiently large $T$ (one can see, making the integral in Eq. (%
\ref{tm_tc}) dimensionless and neglecting $\sqrt{\alpha /T_{CO}}%
Q_{x(y)},\;\mu /T_{CO}$, that it decreases as $\sim 1/\sqrt{T_{CO}}$).

One can see now that the highest value of $T_{CO}$ will be obtained for $%
Q_{y},\;Q_{x}$ maximizing the L.H.S . In the case under consideration, it
implies that both $f(Q_{x},T_{CO})$ and $f(Q_{y},T_{CO})$ should be maximal.

It follows then that the leading instability corresponds to $%
Q_{y}^{max}=Q_{x}^{max}$ and we come to the conclusion that for finite $%
Q_{x,y}^{max}\neq 0$ the diagonal orientation of the CDW wave vector is most
favorable. This correlates with the results of the previous study\cite%
{MetSach,Efetov2013,Sach2013}.

\begin{figure}[h]
\centering
\includegraphics[width=0.8\linewidth]{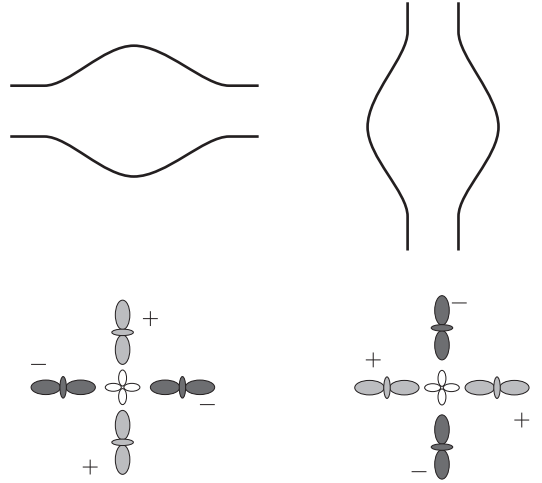}
\caption{Pictorial representation of two possible shapes of the Fermi
surface below the Pomeranchuk transition and the corresponding
intra-unit-cell charge redistributions.}
\label{figpom}
\end{figure}

However, these simple arguments do not exclude an order parameter with $%
Q_{y}^{max}=Q_{x}^{max}=0$. Of course, such an order parameter would no
longer correspond to a CDW, as is evident from (\ref{tm_h}). Instead, the
resulting phase would be characterized by a C$_{4}$-symmetry breaking
deformation of the Fermi surface known in the literature as d-wave
Pomeranchuk instability\cite{yamase2005}. As follows from Eq. (\ref{ZRSbond}%
), such a deformation leads to a redistribution of the charge density
between $p_{x}$ and $p_{y}$ oxygen orbitals (see Fig. \ref{figpom}). An
important property of this state is that it does not open a gap in the
antinodal regions. As will be shown in Section \ref{CDW}, this leaves the
antinodal regions susceptible to further instabilities.

The main result of this Section is that the state with $\mathbf{Q}=0$ is
indeed possible in the model considered here in a certain region of
parameters. To distinguish the order parameter for this state from the one $%
W $ for the conventional CDW we denote it as $P$ and demonstrate that its
finite values are really possible in the model considered. As we have
understood, most favorable should be the state with $\mathbf{Q=}\left(
Q,Q\right) $ maximizing the L.H.S. of Eq. (\ref{tm_tc}).

In other words, we have to find the maximum of the integral $I\left(
Q\right) ,$
\begin{equation}
I\left( Q\right) =\int \frac{dp}{2\pi }\frac{\tanh {\frac{\alpha
(p+Q/2)^{2}-\mu }{2T_{CO}}}-\tanh {\frac{\alpha (p-Q/2)^{2}-\mu }{2T_{CO}}}}{%
2\alpha pQ},  \label{k11}
\end{equation}%
as a function of $Q.$ Writing Eq. (\ref{k11}) we have used Eq. (\ref{tm_disp}%
) and therefore the integrand does not contain the orthogonal momentum $%
p_{\perp }$. Integration over this momentum is replaced by the
multiplication by a constant $\Lambda /2\pi $, where $\Lambda \ll a_{0}^{-1}$
is the size of the relevant antinodal region in the momentum space. The
remaining integral over the momentum $p$ converges and we can extend the
integration limits in the integral $I\left( Q\right) ,$ Eq. (\ref{k11}), to
infinity. This justifies our assumption that the order parameter does not
depend on the momentum. Changing the variables of the integration in Eq. (%
\ref{k11}) to $x=p\sqrt{\alpha /2T_{CO}}$ we reduce the integral $I\left(
Q\right) $ to the form
\begin{equation}
I\left( Q\right) =\frac{\bar{I}\left( q\right) }{8\pi \sqrt{2T_{CO}\alpha }}
\label{k11a}
\end{equation}%
with
\begin{eqnarray}
\bar{I}\left( q\right) &=&\int_{-\infty }^{\infty }\Big[\tanh {\left(
(x+q)^{2}-\frac{\mu }{2T_{CO}}\right) }  \label{tm_susc} \\
&&-\tanh {\left( (x-q)^{2}-\frac{\mu }{2T_{CO}}\right) }\Big]\frac{dx}{qx}
\notag
\end{eqnarray}%
where $q=\sqrt{\frac{\alpha }{2T_{CO}}}Q/2$.

One can clearly see from Eq. (\ref{tm_susc}) that the position of the
maximum is governed by the dimensionless parameter
\begin{equation}
\kappa =\frac{\mu }{2T_{CO}},  \label{k11b}
\end{equation}%
where $T_{CO}$ is an increasing function of $\lambda _{0}$. Numerical
integration in Eq. (\ref{tm_susc}) shows that there exists a critical value $%
\kappa _{cr}$ of the parameter $\kappa $ when the maximum shifts from finite
$q$ to $q=0$. This value can be found analytically by expanding the function
$\bar{I}\left( q\right) ,$ Eq. (\ref{tm_susc}), in $q$. The expansion can be
written as
\begin{equation}
\bar{I}\left( q\right) =\bar{I}\left( 0\right) -b\left( \kappa \right) q^{2},
\label{k12}
\end{equation}%
where%
\begin{equation}
\bar{I}\left( 0\right) =\int_{0}^{\infty }\frac{8}{\cosh ^{2}\left(
x^{2}-\kappa \right) }dx  \label{k12a}
\end{equation}%
and \cite{fexpand}

\begin{equation}
b\left( \kappa \right) =\frac{16}{3}\int_{0}^{\infty }\frac{\sinh
(x^{2}-\kappa )}{\cosh ^{3}(x^{2}-\kappa )}dx.  \label{k13}
\end{equation}

Eq. (\ref{k12a}) shows that $\bar{I}\left( 0\right) >0$ for any $\kappa $.
The dependence of $b\left( \kappa \right) $ on $\kappa $ is more
interesting. Numerical evaluation of the integral $b\left( \kappa \right) $
leads to the result that $b\left( \kappa \right) >0$ for $\kappa <\kappa
_{cr}$ and $b\left( \kappa \right) <0$ for $\kappa >\kappa _{cr},$ where
\begin{equation}
\kappa _{cr}=0.55.  \label{k13a}
\end{equation}

Negative values $b\left( \kappa \right) $ mean that the maximum of L.H.S. of
Eq. (\ref{tm_tc}) cannot be located at $Q=0$ and finite $Q$ are more
favorable. This corresponds to the results of Refs. \cite%
{MetSach,Efetov2013,Sach2013} obtained in the limit $T_{CO}\ll \mu $.
Positive values of $b\left( \kappa \right) $ signal that the charge order
with diagonal modulations does not appear and one comes to the state with $%
Q=0$. So, this state can show up when its transition temperature is higher
than the distance $\mu _{0}$ of the Fermi energy from the saddle point of
the spectrum.

It is interesting to note that the same value of $\kappa =\kappa _{cr}$
leads to the equality
\begin{equation}
\frac{df(0,\kappa _{cr})}{d\mu }=0  \label{k14}
\end{equation}%
for the function $f\left( Q,\kappa \right) $ introduced below Eq. (\ref%
{tm_tc}).

This derivative is negative for $\kappa >\kappa _{cr}$ and positive for $%
\kappa <\kappa _{cr}$. This implies that, provided the Pomeranchuk
instability is the leading one, an increase of the hole doping will result
in growing L.H.S. of (\ref{tm_tc}) and, hence, \textit{decreasing} the
Pomeranchuk transition temperature $T_{Pom}$.

It is useful to obtain analytical expressions for the transition temperature
$T_{pom}$ and the value of the order parameter $P(0)$ at zero temperature.
Taking $Q=0$ in Eq. (\ref{tm_tc}) and introducing the dimensionless units in
the integral one obtains
\begin{equation}
T_{pom}=\frac{1}{8\alpha }\left( \frac{\lambda _{0}\Lambda }{4\pi ^{2}}%
\right) ^{2}\left[ \int dx\frac{1}{\cosh ^{2}(x^{2}-\mu /2T_{pom})}\right]
^{2}  \label{tm_tpom1}
\end{equation}%
The integral in Eq. (\ref{tm_tpom1}) is a slow function of $\mu /2T_{pom}$
when $\mu /2T_{pom}\sim 1$ and is approximately equal to $2$. Then, one has
finally
\begin{equation}
T_{pom}\approx \frac{1}{2\alpha }\left( \frac{\lambda _{0}\Lambda }{4\pi ^{2}%
}\right) ^{2}  \label{tm_tpom2}
\end{equation}%
Note that the critical temperature $T_{pom}$ is proportional to the square
of the coupling constant $\lambda _{0}$ and, in contrast to BCS-like
formulas, can be quite high even for comparatively small values of $\lambda
_{0}.$

To find $P(0)$ one has to first derive a self-consistency equation from Eq. (%
\ref{tm_h}) by taking $Q=0$. This leads to the following equation
\begin{equation}
\frac{1}{2}\sum_{\mathbf{p}}\left[ \tanh {\frac{\varepsilon _{p}^{1}+P(T)}{2T%
}}-\tanh {\frac{\varepsilon _{p}^{2}-P(T)}{2T}}\right] =\frac{2P(T)}{\lambda
_{0}}.  \label{tm_Meq}
\end{equation}%
In the limit $T\rightarrow 0$ the hyperbolic tangent can be replaced by the
sign function, and the integration over the momenta is performed assuming
that $\left\vert P(0)\right\vert >\mu $. This leads to the following
equation
\begin{equation}
\frac{\left\vert P(0)\right\vert }{\lambda _{0}}=\frac{\Lambda }{4\pi ^{2}}%
\sqrt{\frac{\mu +\left\vert P(0)\right\vert }{\alpha }}  \label{k30}
\end{equation}%
The solution of the resulting quadratic equation minimizing the free energy
can be written as
\begin{equation}
\left\vert P(0)\right\vert \approx T_{Pom}+\sqrt{T_{Pom}^{2}+2\mu T_{Pom}}%
>\mu ,  \label{tm_m0}
\end{equation}%
where we have used Eq. (\ref{tm_tpom2}) for $T_{Pom}$. The inequality (\ref%
{tm_m0}) follows from the condition $\kappa <0.55$ that guarantees that we
are in the state with $Q=0$ and justifies the assumptions made when
calculating the integral in Eq. (\ref{tm_Meq}).

The non-zero values of the order parameter $P\left( 0\right) $ do not lead
to a gap in the fermionic spectrum but the Fermi surface gets reconstructed
and acquires a shape like one of those represented in Fig. \ref{figpom}.
This leads to the breaking of the C$_{4}$ symmetry of the original charge
distribution and to opposite excess charges located on $x$ and $y$ orbitals
of the $O$-atoms. It is important to notice that the state is degenerate
because Eq. (\ref{tm_Meq}) allows both $P\left( T\right) $ and $-P\left(
T\right) $ solutions. As a result, two different configurations of the Fermi
surface are possible (one- dimensional anisotropy along either $x$ or $y$
axis).

The results of this Subsection obtained in the framework of the simplified
model demonstrate that the $\mathbf{Q}=0$ charge modulation can indeed occur
in the model specified by the Lagrangian $L_{\mathrm{MF}},$ Eq. (\ref{tm_h}%
). In the next Subsection we consider a more realistic model of fermions
interacting via antiferromagnetic paramagnons and come to similar
conclusions also within that model.

\subsection{\label{PomSF} Spin-Fermion Model.}

We consider the same two regions of the Fermi surface as in Fig. \ref{fig2},
with the single-particle spectrum given by Eq. (\ref{tm_disp}). The
interaction is mediated by critical antiferromagnetic paramagnons as
specified in Eqs. (\ref{b1}-\ref{c6}). Limiting ourselves by consideration
of the regions 1 and 2 as in Fig. \ref{fig2} we reduce the model with the
general action $S$, Eq. (\ref{b1}), to a model with the Lagrangian $L_{%
\mathrm{SF}}$
\begin{equation}
\begin{gathered} L_{\mathrm{SF}} =\sum_{\mathbf{p,}\nu =1,2} \chi
_{\mathbf{p}}^{\nu \dagger } \left[\partial _{\tau }+\varepsilon _{\nu
}\left(\mathbf{p}\right) \right] \chi _{\mathbf{p}}^{\nu }+ \\
+\sum_{q}\vec{\varphi}_{-\mathbf{q}}(-v_{s}^{-2}\partial _{\tau }^{2}+
\mathbf{q}^{2}+a)\vec{\varphi}_{\mathbf{q}} \\ +\lambda \sum_{\mathbf{p,q}}\left[ \chi _{\mathbf{p+q}}^{1\dagger }
\vec{\varphi}_{\mathbf{q}}\vec{\sigma}\chi _{\mathbf{p}}^{2}+\chi _{\mathbf{
p+q}}^{2\dagger }\vec{\varphi}_{\mathbf{q}}\vec{\sigma}\chi
_{\mathbf{p}}^{1} \right] , \label{sf_h} \end{gathered}
\end{equation}%
Writing Eq. (\ref{sf_h}) we have omitted the Coulomb interaction. Its effect
will be taken into account by assuming the d-form factor symmetry of the charge
configurations. In addition, the presence of the Coulomb interaction is
important to reduce the superconducting critical temperature, such that the
Pomeranchuk transition temperature $T_{Pom}$ is the highest critical
temperature in the model.

\subsubsection{Normal state properties.}

\begin{figure}[h]
\includegraphics[trim = 100 150 0 0,width=0.8\linewidth]{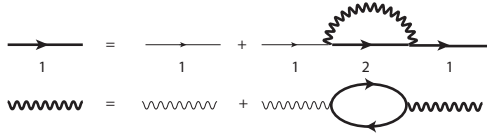}
\caption{Feynman diagrams for fermionic and bosonic propagators illustrating
the approximations used.}
\label{fig3}
\end{figure}

First, we study the normal state (high temperature) properties of this model
because they are different from those obtained in standard considerations in
the vicinity of the Fermi surface. The Green's functions for fermions and
paramagnons have the form
\begin{gather}
G_{\alpha \beta }^{\nu }(i\varepsilon _{n},\mathbf{p})=\frac{\delta _{\alpha
\beta }}{i\varepsilon _{n}-\varepsilon _{\nu }(\mathbf{p})-\Sigma _{\nu
}(i\varepsilon _{n},\mathbf{p})},  \label{k15} \\
D_{mm^{\prime }}(i\omega _{n},\mathbf{q})=-\frac{\delta _{mm^{\prime }}}{%
\left( \omega _{n}/v_{s}\right) ^{2}+\mathbf{q}^{2}+a+\Pi (i\omega _{n},%
\mathbf{q})}.  \notag
\end{gather}%
We calculate the self-energy $\Sigma $ and polarization operator $\Pi $
using the same self-consistent approximation as in Ref. %
\onlinecite{Efetov2013} represented by diagrams in Fig. \ref{fig3}.

The self-consistent approximation used in Ref. \onlinecite{Efetov2013} was
justified by introducing an artificial small angle $\delta $ between the hot
spot Fermi velocities but this is impossible for the present consideration
of the antinodal regions. Fortunately, one can introduce another small
parameter justifying the approximation which turns out to be quite
realistic. At the moment, we neglect the momentum dependence of $\Sigma $
and $\Pi $ and justify this approximation later.

Introducing notations
\begin{eqnarray}
if(\varepsilon _{n}) &\equiv &i\varepsilon _{n}-\Sigma (\varepsilon _{n})
\label{k15a} \\
\left( \lambda /v_{s}\right) ^{2}\Omega (\omega _{n}) &\equiv &\left( \omega
_{n}/v_{s}\right) ^{2}+\Pi (\omega _{n}),  \label{k15b}
\end{eqnarray}%
one obtains

\begin{eqnarray}
&&f^{1(2)}(\varepsilon _{n})-\varepsilon _{n}  \label{k16} \\
&=&-3i\lambda ^{2}T\sum_{\omega _{m},\mathbf{q}}D(i\omega _{m},\mathbf{q}%
)G^{2(1)}(\varepsilon _{n}-\omega _{m},\mathbf{p-q}),  \notag \\
&&\left( \lambda /v_{s}\right) ^{2}\Omega (\omega _{n})-\left( \omega
_{n}/v_{s}\right) ^{2}  \notag \\
&=&2\lambda ^{2}T\sum_{\varepsilon _{n},\mathbf{p}}\Big[G^{1}(\mathbf{p+q}%
,\varepsilon _{n}+\omega _{n})G^{2}(\mathbf{p},\varepsilon _{n})  \label{k18}
\\
&&+G^{2}(\mathbf{p+q},\varepsilon _{n}+\omega _{n})G^{1}(\mathbf{p}%
,\varepsilon _{n})\Big].  \notag
\end{eqnarray}%
Let us now perform the integration over the momenta. First, we calculate the
integral in the electron self-energy
\begin{eqnarray}
&&\int \frac{-1}{\lambda ^{2}\Omega (\varepsilon _{n}^{\prime }-\varepsilon
_{n})+(\mathbf{p-p^{\prime }})^{2}+a}  \notag \\
&&\frac{1}{if(\varepsilon _{n}^{\prime })-\alpha p_{l}^{\prime 2}+\mu }\frac{%
d^{2}p^{\prime }}{(2\pi \hbar )^{2}},  \label{k19}
\end{eqnarray}%
where $l$ stands for $x$ or $y$. Provided the fermionic propagator is more
\textquotedblleft sharp\textquotedblright\ in the momentum space then the
bosonic one, one can perform the momentum integration for the propagators
independently, neglecting the term $\left( p-p^{\prime }\right) _{l}^{2}$ in
the bosonic propagator. Estimating the \textquotedblleft
width\textquotedblright\ of $G$ as $\sqrt{\mu /\alpha }$ and that of $D$ as $%
\sqrt{a}$ we come to the condition $\mu /\alpha \ll a$. In the SF model the
parameter $a$ has the meaning of the inverse square of magnetic correlation
length $\xi ^{-2}$. Therefore, to clarify the physical meaning of this
inequality we rewrite it as
\begin{equation}
\mu v_{s}^{2}/\alpha \ll (v_{s}/\xi )^{2}.  \label{k19a}
\end{equation}

This can be a reasonable assumption, especially taking into account that $%
\xi $ cannot become infinitely large at finite temperatures (see SI of Ref. %
\onlinecite{Efetov2013}). At the same time, we expect the conclusions of our
study to be applicable at least qualitatively even if the inequality (\ref%
{k19a}) does not hold. Performing the integration one obtains for the
integral (\ref{k19})
\begin{eqnarray}
&&\frac{1}{{(2\pi )^{2}}}\frac{\pi ^{2}}{\sqrt{\alpha }}  \label{k20} \\
&&\times \sum_{\varepsilon _{n}^{\prime }}\frac{1}{\sqrt{\left( \lambda
/v_{s}\right) ^{2}\Omega (\varepsilon _{n}-\varepsilon _{n}^{\prime })+a}}%
\frac{\mathrm{sgn}(\mathrm{Re}[f(\varepsilon _{n}^{\prime })])}{\sqrt{%
if(\varepsilon _{n}^{\prime })+\mu }},  \notag
\end{eqnarray}%
where $f(\varepsilon _{n})\equiv f^{1}(\varepsilon _{n})=f^{2}(\varepsilon
_{n})$. This expression manifestly does not depend on momentum $\mathbf{p}$.

The integral over the momentum in the polarization operator $\Pi $ reads
\begin{eqnarray}
&&\int \frac{d^{2}p}{(2\pi )^{2}}\frac{1}{if^{1}(\varepsilon _{n}+\omega
_{n})-\alpha (p_{1}+q_{1})^{2}+\mu }  \notag \\
&&\times \frac{1}{if^{2}(\varepsilon _{n})-\alpha p_{2}^{2}+\mu }
\label{k21}
\end{eqnarray}

The momenta in the fermion propagators are independent and therefore the
result of the integration depends only on the incoming bosonic frequency.
Then, the integral (\ref{k21}) equals
\begin{equation}
\frac{1}{{(2\pi )^{2}}}\frac{\pi ^{2}}{\alpha }\sum_{\varepsilon _{n}}\frac{%
\mathrm{sgn}(\mathrm{Re}[f(\varepsilon _{n})])}{\sqrt{if(\varepsilon
_{n})+\mu }}\frac{\mathrm{sgn}(\mathrm{Re}[f(\varepsilon _{n}+\omega _{n})])%
}{\sqrt{if(\varepsilon _{n}+\omega _{n})+\mu }}.  \label{k22}
\end{equation}%
Introducing an energy scale
\begin{equation}
\Gamma =\left( \frac{\lambda ^{2}v_{s}}{\sqrt{\alpha }\hbar ^{2}}\right)
^{2/3}  \label{k23}
\end{equation}%
and dimensionless variables $\bar{f}=f/\Gamma $, $\bar{\mu}=\mu /\Gamma ,$ $%
\bar{\omega}=\omega /\Gamma $, $\bar{\Omega}=\lambda ^{2}\Omega /\Gamma
^{2}, $ and $\bar{a}=a\left( v_{s}/\Gamma \right) ^{2}$ we can write
equations corresponding to Fig. \ref{fig3} in a dimensionless form
\begin{eqnarray}
&&\bar{f}(\bar{\varepsilon}_{n})-\bar{\varepsilon}_{n}  \label{k24} \\
&=&0.75\bar{T}\sum_{\bar{\varepsilon}_{n}^{\prime }}\frac{1}{\sqrt{\bar{%
\Omega}(\bar{\varepsilon}_{n}-\bar{\varepsilon}_{n}^{\prime })+\bar{a}}}%
\frac{\mathrm{sgn}(\mathrm{Re}[f(\varepsilon _{n}^{\prime })])}{\sqrt{i\bar{f%
}(\bar{\varepsilon}_{n}^{\prime })+\bar{\mu}}},  \notag \\
&&\bar{\Omega}(\bar{\omega}_{n})-\bar{\omega}_{n}^{2}=  \label{k25} \\
&=&-\bar{T}\sqrt{\frac{v_{s}^{2}/\alpha }{\Gamma }}\sum_{\varepsilon _{n}}%
\frac{\mathrm{sgn}(\mathrm{Re}[f(\varepsilon _{n})])}{\sqrt{i\bar{f}(\bar{%
\varepsilon}_{n})+\bar{\mu}}}\frac{\mathrm{sgn}(\mathrm{Re}[f(\varepsilon
_{n}+\omega _{n})])}{\sqrt{i\bar{f}(\bar{\varepsilon}_{n}+\bar{\omega}_{n})+%
\bar{\mu}}}.  \notag
\end{eqnarray}%
One can see that there are three dimensionless parameters $a v_s^2/\Gamma^2$, $%
\mu /\Gamma $ and $\sqrt{v_{s}^{2}/\alpha \Gamma }$ that determine the
behavior of the system. The last parameter is especially important because
it enters the polarization operator but not the fermionic self-energy thus
distinguishing between them. One can also check that the same parameter
enters the renormalization of the vertex part because it contains an
integral over two electron Green's functions as in the polarization
operator. Therefore, in the limit $\sqrt{v_{s}^{2}/\alpha \Gamma }\ll 1$,
one comes to a conclusion that the vertex corrections can be neglected and
the polarization operator might be important only at very low Matsubara
frequencies due to its linear dependence on $\bar{\omega}_{n}$.

To estimate the energy scales $\mu $ and $v_{s}^{2}/\alpha $ we use
experimental data for cuprates. From ARPES data on Bi-2201 presented in
Refs. \onlinecite{Bi2201ARPES-1,Bi2201ARPES-2} we deduce $\mu =|\varepsilon
(\pi ,0)-E_{F}|=25$ meV and $\alpha =\mu /p_{F}^{2}\approx 4.7\cdot 10^{3}$
meV\AA $^{2}$ (lattice constant is $5.44$ \AA ). As there are no inelastic
neutron scattering data available for Bi-2201, we will use the value of $%
v_{s}=200$ meV\AA\ for Bi-2212 from Ref. \onlinecite{prb76214512}. Then, for
$v_{s}^{2}/\alpha $ we obtain the value $\approx 9$ meV. As will be shown
later, our scenario works well for $\bar{\mu}<0.1$ and thus, taking small
values of $\sqrt{v_{s}^{2}/\alpha \Gamma }$ is reasonable.

We can also estimate the region of the magnetic correlation lengths $\xi $
where the momentum-integrated equations are quantitatively correct as $\xi
<13\mathring{A}$. This corresponds to the correlation lengths of the size of
several unit cells. As we consider relatively high temperatures $T\sim
T^{\ast }$, the critical correlation length $\xi $ does not need to be very
large in our theory. In what follows we will present the results of
calculations for $\sqrt{v_{s}^{2}/\alpha \Gamma }=0.5,\;0.1$ and $\bar{a}%
=0.05$.

We have solved the equations (\ref{k24}, \ref{k25}) numerically by iterating
them until the convergence is achieved. Previous treatments restricted to
the vicinities of the hotspots have found $\bar{f}(\bar{\varepsilon})$ to be
purely real. This corresponds to changing the fermionic dispersion $%
i\varepsilon _{n}\rightarrow if(\varepsilon _{n})$. In the present case we
have found that the solution for $\bar{f}(\bar{\varepsilon})$ contains both
real and imaginary parts. The imaginary part of $\bar{f}(\bar{\varepsilon})$
consists of two parts: a temperature-dependent constant $c(\bar{T})$, and a
temperature-independent function of Matsubara frequencies $b(\bar{\varepsilon%
}_{n})$. The constant part enters the fermionic propagator as a
renormalization of $\mu $. As we have considered the problem only in the
antinodal regions, this could mean two things, namely, a
temperature-dependent shift of the chemical potential of the system or a
deformation of the Fermi surface. The latter effect would be possible if
this constant was momentum dependent outside the regions considered here
where our treatment is not applicable. However, as there is no experimental
evidence for such a temperature-dependent deformation, we assume that $c(%
\bar{T})$ can be absorbed into the chemical potential which is fixed by the
total number of particles and therefore can be considered constant at $T\ll
E_{F}$.

\begin{figure}[h]
\includegraphics[width=\linewidth]{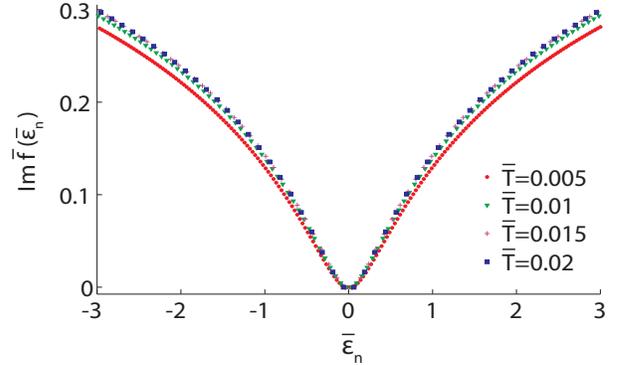}
\caption{$\mathrm{Im}\bar{f}$ as a function of the reduced Matsubara frequency $\bar{\varepsilon}_n=\pi\bar{T}(2n+1)$ for different temperatures. $\bar{a}=0.05$, $%
\protect\sqrt{\frac{v_{s}^{2}/\protect\alpha }{\Gamma }}=0.5$, $\bar{\protect%
\mu}=0.05$.}
\label{fig4}
\end{figure}

The frequency-dependent part of $\mathrm{Im}[\bar{f}(\bar{\varepsilon})]$ is
presented in Fig. \ref{fig4} for several temperatures.
One can see that the function $b$ is clearly temperature-independent. To
understand the physical meaning of this contribution one can perform the
analytical continuation of the resulting self-energy to real frequencies $%
i\varepsilon _{n}\rightarrow \varepsilon $. At low frequencies $b(\bar{%
\varepsilon}_{n})\approx \gamma |\varepsilon _{n}|$ with $\gamma >0$. One
obtains then $\delta \Sigma (i\varepsilon _{n})=\gamma |\varepsilon
_{n}|\rightarrow \delta \Sigma (\varepsilon )=-i\gamma \varepsilon $. Thus,
the physical meaning of this contribution is a quasiparticle damping. Note
that the damping is linear in fermionic frequency in contrast to the usual $%
\varepsilon ^{2}$ Fermi liquid dependence. This is in accord with ARPES
studies of the normal state that show that the quasiparticles in the
antinodal portions of the Fermi surface are strongly damped.

The results for $\Omega (\omega _{n})$ stay in line with the previous
treatment \cite{Efetov2013} of the spin-fermion model. The imaginary part of
this function has been found in all cases to be negligibly small (at the
level of machine precision). The real part exhibits the linear Landau
damping behavior at small frequencies. The constant part $\Omega (0)$ does
not diverge unlike the previous treatment and therefore one can study the
temperature dependence of the bosonic mass $a(T)=a+\left( \lambda
/v_{s}\right) ^{2}\Omega (0,T)$. However, our calculations have shown that
this dependence is weak (at most, $10\%$ difference in the temperature
region of interest) and so we will keep the bosonic mass constant to
simplify calculations.

\subsubsection{Pomeranchuk order.}

Now we will present numerical data for the emerging charge orders. First, we
will compare the critical temperatures $T_{Pom}$ for the Fermi surface
deformation and $T_{diag}$ for the onset of the diagonal modulation. One can
argue in a similar way as in Subsection \ref{Pomsimpl}, that there exists a
critical value of $\bar{\mu}$ below which the Pomeranchuk instability
becomes the leading one (see Appendix \ref{app2}). On the other hand, for
large $\bar{\mu}$ one may linearize the spectrum, which clearly leads to the
diagonal CDW state \cite{Efetov2013,Sach2013}. One can investigate the
transition from one phase to the other in more detail solving mean field
equations numerically. We do not consider now the superconducting phase
assuming that it has been suppressed due to the Coulomb interaction.

The equation for the Pomeranchuk d-wave symmetric order parameter $%
P(\varepsilon ,\mathbf{p})$ can be written in the form
\begin{eqnarray}
&&P\left( \varepsilon ,\mathbf{p}\right)  \label{a19} \\
&=&-\frac{T}{2\left( 2\pi \right) ^{2}}\sum_{\varepsilon ^{\prime },\nu
=1,2}\int \left[ 3\lambda ^{2}D\left( \varepsilon -\varepsilon ^{\prime },%
\mathbf{p-p}^{\prime }\right) +V_{\mathrm{c}}\left( \mathbf{p}-\mathbf{p}%
^{\prime }\right) \right]  \notag \\
&&\times \left( -1\right) ^{\nu -1}\left[ \left( if\left( \varepsilon
^{\prime }\right) -\varepsilon _{\nu }\left( \mathbf{p}^{\prime }\right)
+\mu +\left( -1\right) ^{\nu }P\left( \varepsilon ^{\prime },\mathbf{p}%
^{\prime }\right) \right) \right] ^{-1}d\mathbf{p}^{\prime },  \notag
\end{eqnarray}%
where the momentum integration is performed over the antinodal region.

The critical temperature $T_{Pom}$ can be found linearizing this equation in
$P$
\begin{eqnarray}
&&P\left( \varepsilon ,\mathbf{p}\right) =\frac{T}{\left( 2\pi \right) ^{2}}%
\sum_{\varepsilon ^{\prime }}\int d\mathbf{p}^{\prime }P\left( \varepsilon
^{\prime },\mathbf{p}^{\prime }\right)  \label{e13} \\
&&\times \frac{\left[ 3\lambda ^{2}D\left( \varepsilon -\varepsilon ^{\prime
},\mathbf{p-p}^{\prime }\right) +V_{\mathrm{c}}\left( \mathbf{p}-\mathbf{p}%
^{\prime }\right) \right] }{\left[ f\left( \varepsilon ^{\prime }\right)
+i\left( \varepsilon \left( \mathbf{p}^{\prime }\right) -\mu \right) \right]
^{2}},  \notag
\end{eqnarray}%
where $\varepsilon \left( \mathbf{p}\right) $ is either $\varepsilon
_{1}\left( \mathbf{p}\right) $ or $\varepsilon _{2}\left( \mathbf{p}\right)
. $

Assuming that the order parameter does not depend on $\mathbf{p}$ one can
integrate over momenta and derive the final equations for all $P\left(
\varepsilon \right) $.
\begin{widetext}
\begin{equation}
\begin{gathered}
\bar{f}(\bar{\varepsilon}_{n})-\bar{\varepsilon}_{n}=0.75\bar{T}\sum_{\bar{%
\varepsilon}_{n}^{\prime }}\frac{1}{\sqrt{\bar{\Omega}(\bar{\varepsilon}_{n}-%
\bar{\varepsilon}_{n}^{\prime })+\bar{a}}}\frac{\mathrm{sgn}(\mathrm{Re}[f(%
\bar{\varepsilon}_{n}^{\prime })])}{2}\left[ \frac{1}{\sqrt{i\bar{f}(\bar{%
\varepsilon}_{n}^{\prime })+\bar{\mu}+\bar{P}(\bar{\varepsilon}_{n}^{\prime })}}+%
\frac{1}{\sqrt{i\bar{f}(\bar{\varepsilon}_{n}^{\prime })+\bar{\mu}-\bar{P}(\bar{%
\varepsilon}_{n}^{\prime })}}\right]  \\
\bar{P}(\bar{\varepsilon}_{n})=i\cdot 0.75\bar{T}\sum_{\bar{\varepsilon}%
_{n}^{\prime }}\frac{1}{\sqrt{\bar{\Omega}(\bar{\varepsilon}_{n}-\bar{%
\varepsilon}_{n}^{\prime })+\bar{a}}}\frac{\mathrm{sgn}(\mathrm{Re}[f(\bar{%
\varepsilon}_{n}^{\prime })])}{2}\left[ \frac{1}{\sqrt{i\bar{f}(\bar{%
\varepsilon}_{n}^{\prime })+\bar{\mu}-\bar{P}(\bar{\varepsilon}_{n}^{\prime })}}-%
\frac{1}{\sqrt{i\bar{f}(\bar{\varepsilon}_{n}^{\prime })+\bar{\mu}+\bar{P}(\bar{%
\varepsilon}_{n}^{\prime })}}\right] \\
\\
\bar{\Omega}(\bar{\omega}_{n})-\bar{\omega}_{n}^{2}=-\frac{\bar{T}}{2}\sqrt{\frac{v_{s}^{2}/\alpha }{\Gamma }}
\sum_{\bar{\varepsilon}_{n}}
\left[
\frac
{\mathrm{sgn}(\mathrm{Re}[f(\varepsilon_{n})])}
{\sqrt{i\bar{f}(\bar{\varepsilon}_n)+\bar{\mu}+\bar{P}(\bar{\varepsilon}_n)}}
\frac
{\mathrm{sgn}(\mathrm{Re}[f(\varepsilon _{n}+\omega _{n})])}
{\sqrt{i\bar{f}(\bar{\varepsilon}_{n}+\omega _{n})+\bar{\mu}-\bar{P}(\bar{\varepsilon}_{n}+\omega _{n})}}
+
\right.
\\
\left.
\frac
{\mathrm{sgn}(\mathrm{Re}[f(\varepsilon_{n})])}
{\sqrt{i\bar{f}(\bar{\varepsilon}_{n})+\bar{\mu}-\bar{P}(\bar{\varepsilon}_n)}}
\frac
{\mathrm{sgn}(\mathrm{Re}[f(\varepsilon _{n}+\omega _{n})])}
{\sqrt{i\bar{f}(\bar{\varepsilon}_{n}+\omega _{n})+\bar{\mu}+\bar{P}(\bar{\varepsilon}_{n}+\omega _{n})}}
\right]
.
\end{gathered}
\label{sf_normstate}
\end{equation}
\end{widetext} Equations (\ref{sf_normstate}) are written neglecting the
Coulomb interaction $V_{\mathrm{c}}$ and are used for subsequent numerical
computations.

The sum and the difference of the terms in the square brackets in the first
two equations (\ref{sf_normstate}) guarantee the d-wave symmetry of the
solution for the order parameter $P\left( \bar{\varepsilon}_{n}\right) $.
Equations (\ref{sf_normstate}) have been written for arbitrary temperatures
but, as usual, their linearized version is sufficient for calculation of the
transition temperature $T_{Pom}$.

\begin{figure}[tbp]
\includegraphics[width=\linewidth]{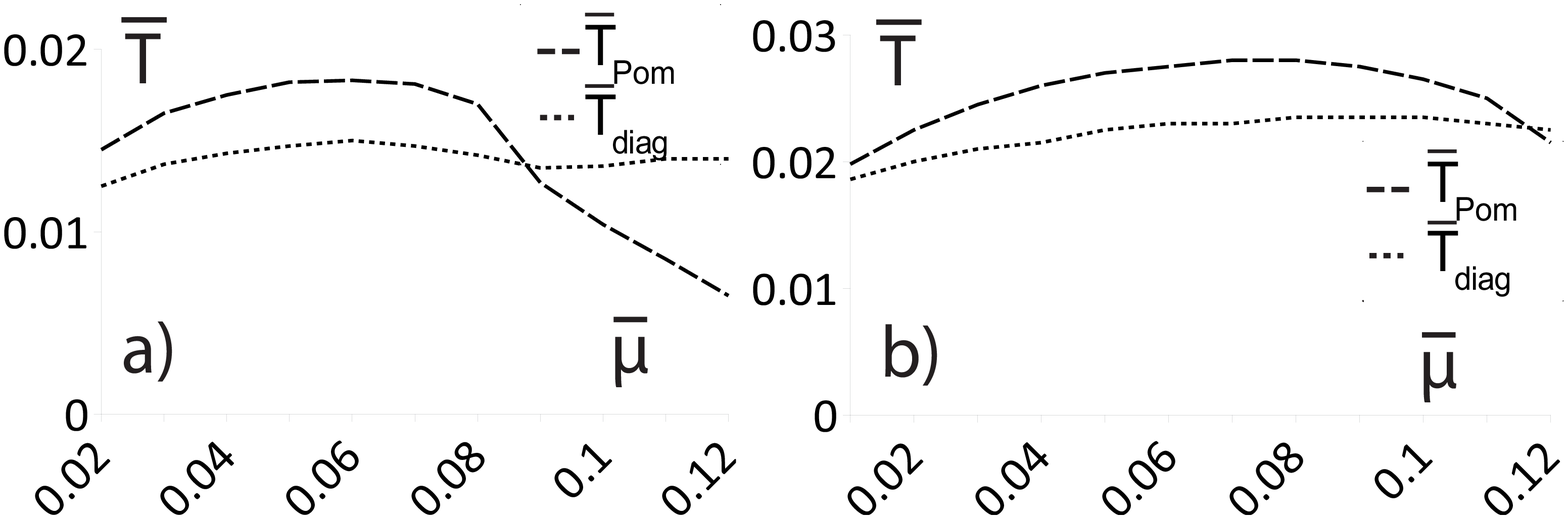}
\caption{$\bar{T}_{Pom}(\bar{\protect\mu})$ (dashed line) and $\bar{T}_{diag}(\bar{\protect%
\mu})$ (dotted line) for $\bar{a}=0.05$, $\protect\sqrt{\frac{v_{s}^{2}/\protect\alpha }{%
\Gamma }}=0.5(a),\;0.1(b)$.}
\label{fig5}
\end{figure}

For diagonal CDW order we will restrict ourselves to calculation of the
transition temperature using the linearized equation
\begin{equation}
\begin{gathered} \bar{W}_{diag}(\bar{\varepsilon}_{n})
=\frac{0.75\bar{T}}{2}\sum_{\varepsilon _{n}^{\prime
}}\frac{\bar{W}_{diag}(\bar{\varepsilon}_{n}^{\prime
})}{\sqrt{\bar{\Omega}(\bar{\varepsilon}_{n}-\bar{\varepsilon _{n}^{\prime
}})+\bar{a}}} \\ \times \frac{\mathrm{sgn}\left(
\mathrm{Re}[f(\bar{\varepsilon}_{n}^{\prime })]\right)
}{\bar{f}(\bar{\varepsilon}_{n}^{\prime
})\sqrt{i\bar{f}(\bar{\varepsilon}_{n}^{\prime })+\bar{\mu}}}.
\label{sf_tdiag} \end{gathered}
\end{equation}%
This equation has been solved numerically by the same iteration procedure as
the one used for solving Eqs. (\ref{k24}, \ref{k25}). The summand in the
R.H.S. has been taken in a slightly non-linear form to improve the
convergence (this obviously does not change the $T_{diag}$ obtained). In
Fig. \ref{fig5} the results for $T_{Pom}(\bar{\mu})$ and $T_{diag}(\bar{\mu}%
) $ are presented for $\bar{a}=0.05$, $\sqrt{v_{s}^{2}/\alpha \Gamma }%
=0.1,\;0.5$.

The crossings of the curves give the critical values $\bar{\mu}_{cr}=0.85$
for the former case and $\bar{\mu}_{cr}=0.12$ for the latter and determine
the region $T_{Pom}>T_{diag}$. In both the cases there is both a critical
value of $\bar{\mu}$ and a value for which $T_{Pom}$ is maximal but, unlike
the simplified model result, they do not coincide. The maximal values of $%
\mu /T_{Pom}$ in the case $T_{Pom}>T_{diag}$ are obtained at $\bar{\mu}_{cr}$
and are $\left( \mu /T_{Pom}\right) _{cr}=5.5$ and $\left( \mu
/T_{Pom}\right) _{cr}=4.4$ for the two cases considered. These values are
considerably larger the value $\approx 1.1$ obtained in the simplified model
from Eqs. (\ref{k11b}, \ref{k13a}). As is shown in Appendix \ref{app2}, this
is a consequence of the renormalization of the fermionic dispersion $\bar{f}(%
\bar{\varepsilon}_{n})$. Assuming that $T_{Pom}$ is of the order of the
pseudogap temperature but higher than the latter we conclude that the most
appropriate values for $\bar{\mu}$ in the moderately underdoped regime
should be around $0.03-0.06$.

This implies that $T_{Pom}$ is of the same order of magnitude as $\mu $
(probably 2-3 times smaller), which makes it quite possible that $%
T_{Pom}>T^{\ast }\geq T_{CDW}$. It is not surprising then that a charge
modulation with a diagonal modulation vector has never been observed.
According to the present results, the Pomeranchuk instability prevails
changing the scenario for formation of the CDW. Formation of the diagonal
modulation of Refs. \onlinecite{MetSach,Efetov2013} requires considerably
higher values of $\mu $ than those observed experimentally for the
hole-doped cuprates.

\begin{figure}[h]
\includegraphics[width=0.8\linewidth]{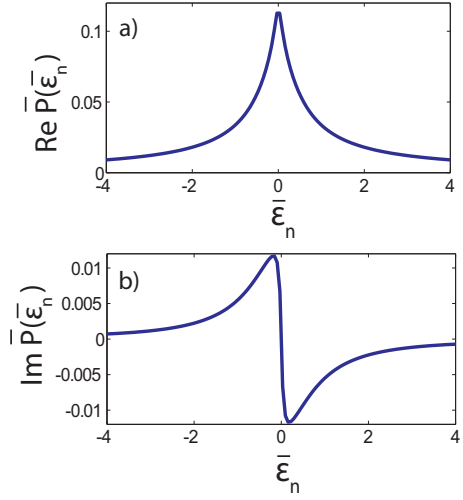}
\caption{Dependence of real (a) and imaginary (b) parts of the Pomeranchuk order parameter $\bar{P}$ on the reduced Matsubara frequency $\bar{\varepsilon}_n=\pi\bar{T}(2n+1)$, $\bar{a}=0.05$, $\bar{\protect\mu}=0.05$, $\protect\sqrt{%
\frac{v_{s}^{2}/\protect\alpha }{\Gamma }}=0.5$, $\bar{T}=0.01$.}
\label{figreim}
\end{figure}

\begin{figure}[h]
\includegraphics[width=0.7\linewidth]{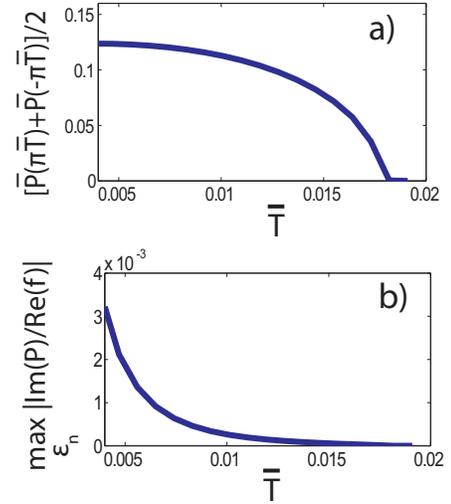}
\caption{Dependence of the Pomeranchuk order parameter $\bar{P}$ on temperature: a) Height of the peak in the real part $0.5(\bar{P}(\protect\pi \bar{T})+\bar{P}(-\protect\pi \bar{T}))$ and b) Relative magnitude of the imaginary part given by $\max_{\varepsilon_n}\left|{\rm Im} P(\varepsilon_n)/{\rm Re} f(\varepsilon_n)\right|$. The model parameters are $\bar{a}=0.05$, $\protect%
\sqrt{\frac{v_{s}^{2}/\protect\alpha }{\Gamma }}=0.5$.}
\label{figpomT}
\end{figure}

Equations (\ref{sf_normstate}) allow one to compute the order parameter $%
P\left( \bar{\varepsilon}\right) $ as a function of the reduced temperature $%
\bar{T}$ and frequency $\bar{\varepsilon}$. The result of the computation is
presented in Figs. \ref{figreim},\ref{figpomT}.

The non-zero values of the Pomeranchuck order parameter result in a
quasi-one-dimensional shape of the Fermi surface like the one represented in
Fig. \ref{figpom}. As a result, the bond correlations $\left\langle \chi
_{i}^{\ast }\chi _{i+a_{x}}\right\rangle $ and $\left\langle \chi _{i}^{\ast
}\chi _{i+a_{y}}\right\rangle $ in the SF model differ from each other.
Using the correspondence between the bond correlations in the SF model and
charges on the $O$ atoms on the $CuO_{2}$ lattice (SI of Ref. %
\onlinecite{Efetov2013} and Appendix \ref{app1a}) we conclude that the
charges on the $O_{x}$ and $O_{y}$ orbitals are different. The resulting
picture is displayed in Fig. \ref{figpom}.

The plot in Fig. \ref{figpomT} and the charge distribution in Fig. \ref%
{figpom} are applicable at all temperatures below $T_{Pom}$ only if there
are no other instabilities in this region. At the same time, the Fermi
surface remains ungapped below $T_{Pom}$ and there is no reason to exclude
additional phase transitions. In the next Section we will demonstrate that a
d-form factor CDW with modulation along the BZ axes and d-wave
superconductivity are indeed possible.

\section{\label{CDW}Charge density wave and superconductivity in the model
with the deformed Fermi surface.}

\subsection{\label{CDWsimpl}CDW in the simplified model.}

As in the previous Section, we consider first the simplified model with the
constant interaction between the regions 1 and 2 in Fig. \ref{fig2}. Now we
analyze d-form factor particle-hole instabilities at temperatures below the
Pomeranchuk transition temperature, $T<$ $T_{Pom}$. As the order parameter $%
P\left( T\right) $ only shifts the \textquotedblleft chemical
potentials\textquotedblright\ $\mu $ in the regions 1 and 2, the equation
for $T_{CO}$ (\ref{tm_tc}) remains intact provided the energy spectra $%
\varepsilon _{p}^{1}$ and $\varepsilon _{p}^{2}$ are modified as
\begin{gather}
\varepsilon _{p}^{1}\rightarrow \alpha p_{x}^{2}-\mu +P(T),  \label{c1} \\
\varepsilon _{p}^{2}\rightarrow \alpha p_{y}^{2}-\mu -P(T).  \notag
\end{gather}

It follows then that the L.H.S. of Eq. (\ref{tm_tc}) is a sum of two
functions that we have already analyzed. However, the control parameters are
different because instead of $\kappa \left( T\right) =\mu /2T$, one has
either $\kappa _{-}=(\mu -P(T))/2T$ in the first region or $\kappa _{+}=(\mu
+P(T))/2T$ in the second one. As $\left\vert P(T)\right\vert $ grows when
decreasing the temperature, it is clear that below a certain temperature $%
T<T_{Pom}$ the susceptibility contribution from the second (from the first,
if $P<0$) region will reach the maximum at a nonzero $\mathbf{Q}$ directed
along $y$($x$). This happens because $\kappa _{+}\left( T\right) $ will
inevitably exceed $\kappa _{cr}=0.55$ with decreasing the temperature. In
the other region the situation is more subtle because, although $\mu -|P(T)|$
clearly decreases, so does the temperature $T$. It is therefore important to
understand if $(\mu -|P(T)|)/2T$ can become larger than the critical value $%
\kappa _{cr}=0.55$ when the maximum starts to move from $\mathbf{Q}=0$ to
finite $\mathbf{Q}$. Let us concentrate on the case $P\left( T\right) >0$.

An useful inequality can be derived introducing a function
\begin{equation}
A\left( T\right) =\left\vert P\left( 0\right) \right\vert \sqrt{1-T/T_{pom}}
\label{c2}
\end{equation}%
The function $A\left( T\right) $ cannot serve as a good approximation to $%
P\left( T\right) $ even in the vicinity of $T_{Pom}$ because the coefficient
in the R.H.S. of Eq. (\ref{c2}) is smaller then the one given by the
solution of mean field equations. For example, for the Ising model, the
order parameter $P\left( T\right) =\sqrt{3}A\left( T\right) $ when $%
T\rightarrow T_{Pom}$. Generally, the inequality
\begin{equation}
P\left( T\right) \geqslant A\left( T\right)  \label{c2a}
\end{equation}%
holds.

The statement (\ref{c2a}) can be proven using the fact that $P\left(
0\right) =A\left( 0\right) $ and $P\left( T_{Pom}\right) =A\left(
T_{Pom}\right) $. As both the functions are monotonously decaying with $T$
and $\left\vert dP\left( T\right) /dT\right\vert <\left\vert dA\left(
T\right) /dT\right\vert $ as $T\rightarrow 0$ (the function $P\left(
T\right) $ approaches $P\left( 0\right) $ exponentially in $1/T$, while $%
A\left( T\right) $ does it linearly in $T$), one comes the inequality (\ref%
{c2a}).

This allows us to write the following inequalities
\begin{gather}
\frac{\mu -P(T)}{2T}\leqslant \frac{\mu -A\left( T\right) }{2T}  \label{c3}
\\
=\frac{\mu T/T_{Pom}+\mu (1-T/T_{pom})-A\left( T\right) }{2T}<\frac{\mu }{%
2T_{Pom}}.  \notag
\end{gather}%
The inequality in the second line of (\ref{c3}) follows immediately from the
inequality (\ref{tm_m0}) and the definition of the function $A\left(
T\right) $, Eq. (\ref{c2}).

Inequalities (\ref{c3}) show that $\kappa _{-}\left( T\right) <\kappa
_{Pom}, $ where $\kappa _{Pom}=\mu /2T_{Pom}.$ As we are below the
Pomeranchuk critical temperature $T_{Pom},$ we have $\kappa _{Pom}<\kappa
_{cr}$ and, hence, $\kappa _{-}\left( T\right) <\kappa _{cr}$. With this
result we come to the conclusion that the contribution of the region with
the Fermi surface shrinking due to the distortion has always the maximum at
the zero vector of modulation.

These simple arguments allow us to guarantee that when decreasing the
temperature the Pomeranchuk transition with $\mathbf{Q}=0$ is followed by a
transition into a d-form factor CDW state with the vector of modulation
directed along the axes of the BZ. For $P\left( T\right) >\mu $ one comes to
the Fermi surface represented in the left part of Fig. \ref{figpom} and the
modulation vector is directed along the y-axis, while for $P\left( T\right)
<-\mu $ the picture should be turned by $90%
{{}^\circ}%
$. The above arguments do not exclude formation of CDW even for $-\mu
<P\left( T\right) <\mu $ but due to the inequality (\ref{tm_m0}) one
inevitably has $\left\vert P\left( T\right) \right\vert >\mu $ at
sufficiently low temperatures.

Let us now estimate the transition temperature for the CDW. We will carry
out the calculations assuming $T_{CDW}\ll T_{pom}$ and checking this
assumption afterwards. In this limit one can approximate the function $P(T)$
by its value $P\left( 0\right) $ at zero temperature, Eq. (\ref{tm_m0}).
Then, the term in Eq. (\ref{tm_tc}) having the maximum at the zero wave
vector is proportional to $\exp {(\mu -|P(0)|)/T}\ll 1$ and therefore can be
neglected. In the other term, one has an \textquotedblleft effective Fermi
energy\textquotedblright\ ${(\mu +|P(0)|)\gg T}$. In this limit the
calculation of the integral in the L.H.S. of Eq. (\ref{tm_tc}) is similar to
a standard calculation of the corresponding integral for a CDW instability
in a system with a nesting and a large Fermi energy. Then, the magnitude of
the wave vector maximizing the term in Eq. (\ref{tm_tc}) is given by
\begin{equation}
Q=2\sqrt{(\mu +\left\vert P(0)\right\vert )/\alpha },  \label{c3a}
\end{equation}%
which corresponds to the vector connecting the nesting points in the
conventional CDW instability. As a result, one comes to the following
equation
\begin{equation}
\frac{\Lambda }{8\pi ^{2}}\int_{-\infty }^{\infty }\frac{\tanh {\frac{\alpha
p^{2}+2\xi \left( p\right) }{2T_{CDW}}}-\tanh {\frac{\alpha p^{2}-2\xi
\left( p\right) }{2T_{CDW}}}}{4p\sqrt{\alpha (\mu +|P(0)|)}}dp=\frac{2}{%
\lambda _{0}},  \label{c4}
\end{equation}%
where $\xi \left( p\right) =p\sqrt{\alpha \left( \mu +P\left( 0\right)
\right) }.$

Changing from the variables $p$ to $\xi \left( p\right) $ one can easily
calculate the integral in Eq. (\ref{c4}) to obtain the critical temperature $%
T_{CDW}$ of the transition to the CDW state

\begin{eqnarray}
T_{CDW} &\approx &\frac{\pi }{4e^{\gamma }}(\mu +|P(0)|)\exp \left( -\frac{%
16\pi ^{2}\sqrt{\alpha (\mu +|M(0)|)}}{\lambda \Lambda }\right)  \notag \\
&\approx &0.44(\mu +|P(0)|)\exp \left( -4\sqrt{\frac{\mu +\left\vert P\left(
0\right) \right\vert }{2T_{pom}}}\right) ,  \label{c5}
\end{eqnarray}%
where $\gamma \approx 0.5772$ is the Euler gamma constant and $T_{Pom}$ is
determined by Eq. (\ref{tm_tpom2}). For $T_{Pom}\sim \mu $ numerical
evaluation in Eq. (\ref{c5}) leads to the estimate $T_{CDW}\sim 0.007\mu \ll
T_{Pom}$ with a CDW wave vector magnitude $2\sqrt{(2+\sqrt{3})\mu /\alpha }%
\approx 1.9Q_{0}$, where $Q_{0}=2\sqrt{\mu /\alpha }$ is the vector
connecting the antinodal points of the FS in case of small curvature
(otherwise it is larger). As $\mu$ decreases with hole doping, so does $Q_0$, reproducing the qualitative behavior seen in the experiments\cite{y123REXS-3,y123XRD-3,bi2201STM-1}. While the resulting modulation wave vector seems
to be in a good agreement with experimental data \cite{bi2201STM-2}, the
large difference between the temperatures $T_{Pom}\sim \mu $ and $T_{CDW}$
clearly contradicts the experimentally observed $T_{CDW}\sim 100$ K.

A possible scenario making $T_{CDW}$ closer to $T_{Pom}$ can be formulated
as follows: as the Pomeranchuk distortion develops, the region where the
Fermi surface shrinks can become nearly nested with a modulation vector
having the same direction as in the other region. In the best case, both
regions are going to have precisely the same nesting wave vector. Then, one
can estimate the transition temperature by taking both contributions in Eq. (%
\ref{tm_tc}) to be the same. This leads to $T_{CDW}\approx 0.44(\mu
+|P(0)|)\exp \left( -\sqrt{2\left( \mu +|P(0)|\right) /T_{Pom}}\right)
\approx 0.078\mu $, which is still too small for a quantitative agreement.
Nevertheless, the qualitative scenario of a CDW transition preempted by a
Fermi surface deformation transition gives a hint to the robustness of the
CDW vector direction in the cuprates. In what follows, we will show that a
more realistic frequency-dependent interaction of the Spin-Fermion model
provides a much better quantitative estimates for $T_{CDW}$ and $Q_{CDW}$
together with a more relaxed constraint on the value of $\mu $.

Up to this point the consideration has been performed on the mean field
level. However, fluctuations and inhomogeneities can manifest themselves in
the proposed mean field scenario. The Pomeranchuk order parameter $P$ breaks
the discrete symmetry and therefore the long-range order is not destroyed
even in the strictly 2D case. Inhomogeneities, however, can be energetically
profitable and proliferate in a form of domains with different signs of the
order parameter. As the sign of $P$ sets the direction of the CDW, different
domains will have CDWs directed in $x$ or $y$ direction depending on the
sign of $P$. This indeed corresponds to recent STM\cite{Bi2212STM-2}, RXS%
\cite{y123REXS-4} and XRD\cite{HgXRD} experiments. Note that this also
provides a mechanism of \textquotedblleft masking\textquotedblright\ the C$%
_{4}$ symmetry breaking on the global scale alternative to the one proposed
in Ref. \onlinecite{yamase2009}. Unlike the Pomeranchuk order, CDW breaks a
continuous translation symmetry and thus the transition should necessarily
be smeared. Moreover, our scenario is fully compatible with the ideas of
Ref. \onlinecite{Efetov2013}, \textit{i.e.} the competing orders, such as
superconductivity or antiferromagnetism at lower dopings can induce an
orderless pseudogap state, while lowering the ordering temperature. Thus,
while it is tempting to assume $T_{CDW}\approx T^{\ast }$ in the presented
scenario, fluctuations can certainly lead to $T_{CDW}<T^{\ast }$, a
situation consistently observed in YBCO\cite{y123XRD-3,y123REXS-3}.
Summarizing, the qualitative conclusions of the simplified model are:

\begin{itemize}
\item Provided the interaction is sufficiently strong with respect to $%
|\varepsilon (0,\pi )-E_{F}|,$ the Pomeranchuk instability is the leading
one in the d-form factor particle-hole channel

\begin{itemize}
\item There are no phase fluctuations for the Pomeranchuk order parameter
and therefore the long-range order is not destroyed in 2D.

\item The order can have domain structure with different domains
accommodating Fermi surface distortion either in $x$- or in $y$-directions.
\end{itemize}

\item At $T_{CDW}<T_{Pom}$ a transition into a d-form factor CDW state
occurs with the CDW modulation vector being directed along one of the BZ
edges.

\begin{itemize}
\item The direction of $\mathbf{Q}_{CDW}$ is determined by the sign of the
Pomeranchuk order parameter. This implies that domains with different signs
of the deformation of the Fermi surface will host CDWs with different
modulation vectors.

\item The magnitude of $\mathbf{Q}_{CDW}$ depends on the magnitude of the
Pomeranchuk order parameter at $T_{CDW}$ and, hence, is not universally
related to $Q_{AN}$ (vector connecting adjacent antinodes) or $Q_{HS}$
(distance between hotspots or tips of the Fermi arcs). Hole doping leads to the decrease of $|\mathbf{Q}_{CDW}|$.
\end{itemize}
\end{itemize}

\subsection{\label{CDWSF}CDW in the Spin-Fermion model.}

Now we consider formation of CDW below $T_{Pom}$ in the SF model. As the
Fermi surface is not $C_{4}$ symmetric, the order parameters for purely
paramagnon interaction do not necessarily satisfy in two regions 1 and 2 the
equality $W_{1}=-W_{2}$ implied in the simplified model of the previous
Subsection. This leads to presence of an on-site modulation (s-form factor
component). However, as has been discussed in Section \ref{Constr}, the
strong on-site Coulomb repulsion suppresses on-site modulations. In
principle, in order to take it into account one should solve the full
equations (\ref{k8}).

Here we will follow a different route. In order to simplify computations,
one can replace the first term in R.H.S. of Eq. (\ref{k8}) by the constraint
\begin{equation}
T\sum_{\varepsilon }\int \rho _{\mathbf{Q}}\left( \varepsilon ,\mathbf{p}%
\right) d^{3}\mathbf{p}=0,  \label{c7}
\end{equation}%
where
\begin{equation}
\rho _{\mathbf{Q}}\left( \varepsilon ,\mathbf{p}\right) =\left\langle \chi
_{\varepsilon ,\mathbf{p+Q/2}}^{\dagger }\chi _{\varepsilon ,\mathbf{p-Q/2}%
}\right\rangle  \label{c8}
\end{equation}%
Equation (\ref{c7}) means that the total modulated charge in the elementary
cell equals zero. In particular, the charge modulation on the $Cu$ atoms
vanishes due to this constraint and the s-component of the charge
distribution does not arise. The $O$ atoms are not explicitly present in the
one-band SF model considered here. According to Appendix \ref{app1a} the
correlations $\left\langle \chi _{\mathbf{r}}^{\dagger }\chi _{\mathbf{r+a}%
}\right\rangle ,$ where $\mathbf{a}$ is the lattice vector, determine
charges on the $O$-atoms located on the bonds of $CuO_{2}$ lattice
connecting the points $\mathbf{r}$ and $\mathbf{r+a}$. Within the
approximations adopted in this paper the order parameter is momentum
independent inside the regions 1 and 2 of Fig. \ref{fig2} and therefore the
charge distribution should have the d-form factor as soon as the
constraint (\ref{c7}) is fulfilled.

For practical calculations the constraint (\ref{c7}) is not very convenient
and we replace it by a stronger one%
\begin{equation}
\rho _{\mathbf{Q}}^{1}\left( \varepsilon \right) +\rho _{\mathbf{Q}%
}^{2}\left( \varepsilon \right) =0.  \label{c9}
\end{equation}%
In Eq. (\ref{c9}) the superscripts relate to the regions 1 and 2 in Fig. \ref%
{fig2} but the dependence of the order parameters on the momentum is
neglected inside these regions. It is easy to see that Eq. (\ref{c9}) leads
as previously to the condition%
\begin{equation}
W_{\mathbf{Q}}^{1}\left( \varepsilon \right) =-W_{\mathbf{Q}}^{2}\left(
\varepsilon \right)  \label{c10}
\end{equation}

Note that provided the order parameter does not depend on frequency, as was
the case in the simplified model, the use of the two constraints, (\ref{c7})
and (\ref{c9}) results in the same equations.

In principle, to find $T_{CDW}$ one should consider the self-consistency
equations for CDW with an arbitrary wave vector $Q$ and then chose $Q$
yielding the largest transition temperature. We will obtain an estimate for $%
T_{CDW}$ taking $Q$ based on the results of the simplified model. $Q$ has
been found to be directed along BZ axis with its magnitude given by Eq. (\ref%
{c3a}). As we assume that $T_{CDW}$ can be close to $T_{Pom},$ we should
accept that $P(T_{CDW})\neq P(0)$. Moreover, in the SF model the order
parameter $P$ depends on the Matsubara frequency. We take these properties
into account by generalizing the expression (\ref{c3a}):
\begin{equation}
Q^{SF}(T)=2\sqrt{(\mu +0.5\left\vert P(-\pi T)+P(\pi T)\right\vert )/\alpha }%
.  \label{sf_q}
\end{equation}%
Using this magnitude of the wave vector and taking Eq. (\ref{c9}) into
account one obtains after momentum integration
\begin{equation}
\begin{gathered}
\bar{W}(\bar{\varepsilon}_{n})=0.75\;i\frac{\bar{T}_{CDW}}{2}\sum_{%
\varepsilon _{n}^{\prime
}}\frac{\mathrm{sgn}(\mathrm{Re}[f(\bar{\varepsilon}_{n}^{\prime
})])}{2\sqrt{\bar{\Omega}(\bar{\varepsilon}_{n}-\bar{\varepsilon
_{n}^{\prime }})+\bar{a}}} \\ \times \left[
\frac{\bar{W}(\bar{\varepsilon}_{n}^{\prime })}{\left[
(i\bar{f}(\bar{\varepsilon}_{n}^{\prime
})+\bar{P}(\bar{\varepsilon}_{n}^{\prime })-\bar{P}(0)\right] g\left(
\bar{\varepsilon}_{n}^{\prime }\right)
}+\frac{\bar{W}(\bar{\varepsilon}_{n}^{\prime })}{g^{3}\left(
\bar{\varepsilon}_{n}^{\prime }\right) }\right] , \end{gathered}  \label{c12}
\end{equation}%
where $g\left( \bar{\varepsilon}_{n}\right) =\sqrt{i\bar{f}(\bar{\varepsilon}%
_{n})+\bar{\mu}+\bar{P}(\bar{\varepsilon}_{n})}$ and $\bar{P}(0)=0.5\left(
\bar{P}(-\pi \bar{T})+\bar{P}(\pi \bar{T})\right) $

The solution for this equations suffers from the same problem as the one for
the simplified model, namely, $T_{CDW}$ is too small. For $\sqrt{%
v_{s}^{2}/\alpha \Gamma }=0.5$ it is at least an order of magnitude smaller
then $\mu $ except for the lowest values of $\mu $ implying $T_{CDW}\sim
10-30$ $K$. Similar to what we have done for the simplified model, we would
like to see if the result changes under the assumption that in the region
where Fermi surface shrinks the nesting emerges. The equation for $T_{CDW}$
is then:
\begin{eqnarray}
&&\bar{W}(\bar{\varepsilon}_{n})=0.75\;i\frac{\bar{T}_{CDW}}{2}%
\sum_{\varepsilon _{n}^{\prime }}\frac{\bar{W}(\bar{\varepsilon}_{n}^{\prime
})}{\sqrt{\bar{\Omega}(\bar{\varepsilon}_{n}-\bar{\varepsilon _{n}^{\prime }}%
)+\bar{a}}}  \notag \\
&&\times \frac{\mathrm{sgn}(\mathrm{Re}[f(\bar{\varepsilon}_{n}^{\prime })])%
}{(\left[ i\bar{f}(\bar{\varepsilon}_{n}^{\prime })+\bar{P}(\bar{\varepsilon}%
_{n}^{\prime })-P(0)\right] g\left( \bar{\varepsilon}^{\prime }\right) }
\label{c11}
\end{eqnarray}%
\begin{figure}[tbp]
\includegraphics[width=\linewidth]{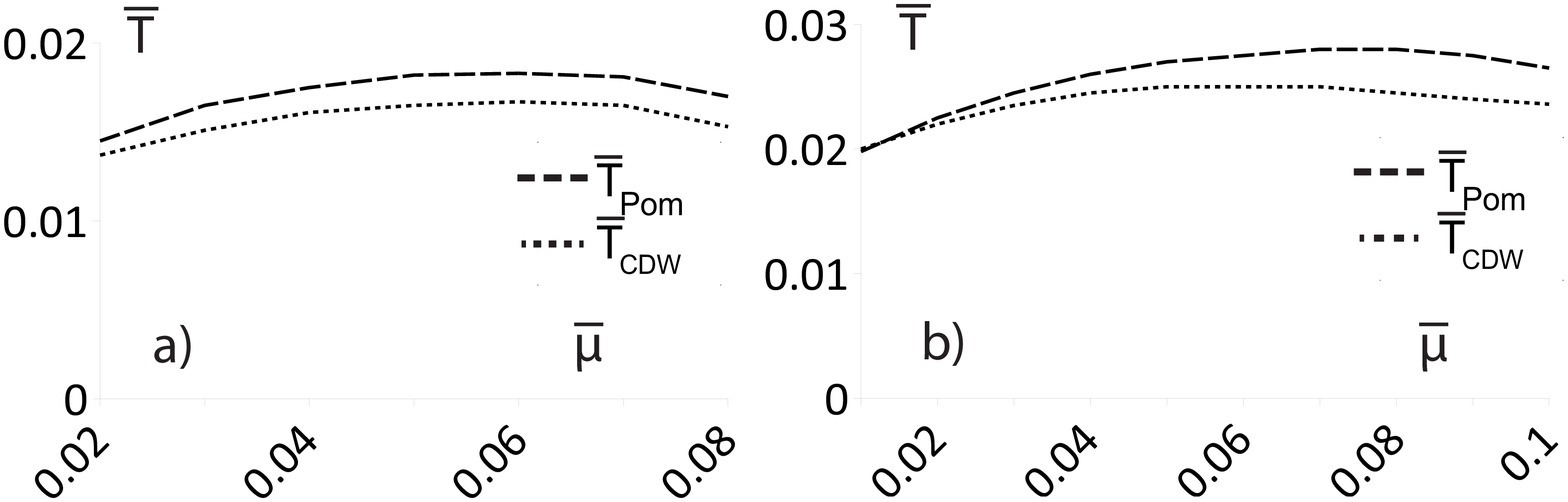}
\caption{$T_{Pom}(\bar{\protect\mu})$ (dashed line) and $\bar{T}_{CDW}(\bar{%
\protect\mu})$ (dotted line) determined from Eq. (\protect\ref{c11}) for $%
\bar{a}=0.05$, $\protect\sqrt{\frac{v_{s}^{2}/\protect\alpha }{\Gamma }}%
=0.5(a),\;0.1(b)$.}
\label{fig6}
\end{figure}
Results of the numerical solution are presented in Fig. \ref{fig6}. One can
see that in this case $\bar{T}_{CDW}$ is very close to $\bar{T}_{Pom}$
making the scenario quantitatively viable. One should ask, however, if $P$
is large enough at $T_{CDW}$ to sufficiently affect the Fermi surface. In
Fig. \ref{fig7} the value of $0.5(P(\pi T)+P(-\pi T))$ at $T_{CDW}$ is
given. One can see that the values are large enough to completely shrink the
Fermi surface in one of the regions making the \textquotedblleft
nesting\textquotedblright\ assumption reasonable. From Fig. \ref{fig7} one
can also estimate using Eq. (\ref{sf_q}) the magnitude of $Q_{CDW}$ that
turns out to be close to $1.5Q_{0}$. This also means that $Q_{CDW}$ decreases with decreasing $\bar{\mu}$, e.g. with hole doping, consistent with the experiments\cite{y123REXS-3,y123XRD-3,bi2201STM-1}.
\begin{figure}[tbp]
\includegraphics[width=\linewidth]{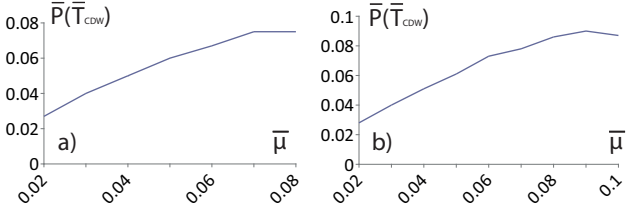}
\caption{Value of the Pomeranchuk order parameter at the CDW transition $0.5(%
\bar{P}(\protect\pi \bar{T})+\bar{P}(-\protect\pi \bar{T}))$ for $\bar{a}%
=0.05$, $\protect\sqrt{\frac{v_{s}^{2}/\protect\alpha }{\Gamma }}%
=0.5(a),\;0.1(b)$.}
\label{fig7}
\end{figure}
Altogether, the results of the spin-fermion model treatment are:

\begin{itemize}
\item Spin-fermion model near the saddle point of the electron spectrum has
a small parameter $v_{s}^{2}/\alpha \Gamma $ justifying an Eliashberg-type
approximation.

\item For the normal state the spin-fermion model near the saddle point
yields a strong (linear) damping of antinodal quasiparticles.

\item The results for the d-form factor charge ordering qualitatively agree
with the simplified model from Section \ref{Pom}.

\item Quantitatively viable results for $T_{CDW}$ can be obtained by taking
into account the emerging nesting in the region with shrinking Fermi surface.
\end{itemize}

\subsection{Superconductivity in the extended SF model.}

Until now we have been considering charge modulations. However,
superconducting phase is very important in cuprates as well as in the model
considered here. It competes with the charge modulation and it is important
to understand when it can win and when it cannot. Adopting Eq. (\ref{k7}) to
the model with the two regions 1 and 2, Fig. \ref{fig2}, we write the
equation for the superconducting order parameter in the form

\begin{eqnarray}
&&\Delta \left( \varepsilon ,\mathbf{p}\right) =\frac{T}{\left( 2\pi \right)
^{2}}\sum_{\varepsilon ^{\prime }}\int d\mathbf{p}^{\prime }\Big[3\lambda
^{2}D\left( \varepsilon -\varepsilon ^{\prime },\mathbf{p-p}^{\prime }\right)
\notag \\
&&-V_{\mathrm{c}}\left( \mathbf{p}-\mathbf{p}^{\prime }\right) \Big]
\label{a49} \\
&&\times \Delta \left( \varepsilon ^{\prime },\mathbf{p}^{\prime }\right) %
\left[ |if\left( \varepsilon ^{\prime }\right) +\varepsilon \left( \mathbf{p}%
^{\prime }\right) -\mu |^{2}+\Delta ^{2}\left( \varepsilon ^{\prime },%
\mathbf{p}^{\prime }\right) \right] ^{-1}.  \notag
\end{eqnarray}%
A non-trivial solution for the superconducting $\Delta \left( \varepsilon ,%
\mathbf{p}\right) $ appears at a critical temperature $T_{c}$ that can be
found linearizing Eq. (\ref{a49}) in $\Delta \left( \varepsilon ,\mathbf{p}%
\right) $
\begin{eqnarray}
&&\Delta \left( \varepsilon ,\mathbf{p}\right) =\frac{T}{\left( 2\pi \right)
^{2}}\sum_{\varepsilon ^{\prime }}\int \Big[3\lambda ^{2}D\left( \varepsilon
-\varepsilon ^{\prime },\mathbf{p-p}^{\prime }\right)  \notag \\
&&-V_{\mathrm{c}}\left( \mathbf{p}-\mathbf{p}^{\prime }\right) \Big]\Delta
\left( \varepsilon ^{\prime },\mathbf{p}^{\prime }\right) |if\left(
\varepsilon ^{\prime }\right) +\varepsilon \left( \mathbf{p}^{\prime
}\right) -\mu |^{-2}d\mathbf{p}^{\prime }.  \notag \\
&&  \label{a51}
\end{eqnarray}%
Equation (\ref{a51}) for the superconducting order parameter $\Delta $
differs from Eq. (\ref{e13}) for the Pomeranchuk order parameter $P$ by the
opposite sign in front of the Coulomb interaction and by the combination $%
|if\left( \varepsilon ^{\prime }\right) +\varepsilon \left( \mathbf{p}%
^{\prime }\right) -\mu |^{-2}$ instead of $\left[ f\left( \varepsilon
^{\prime }\right) +i\left( \varepsilon _{1}\left( \mathbf{p}^{\prime
}\right) -\mu \right) \right] ^{-2}$ in the integrand.

Neglecting the Coulomb interaction and performing the momentum integration
one arrives at:
\begin{gather*}
\bar{\Delta}(\varepsilon )=0.75\bar{T}\sum_{\varepsilon ^{\prime }}\frac{%
\bar{\Delta}(\varepsilon ^{\prime })}{\sqrt{\bar{\Omega}(\varepsilon
-\varepsilon ^{\prime })+\bar{a}}} \\
\frac{1}{\sqrt{\bar{\Delta}^{2}(\varepsilon ^{\prime })+\mathrm{Re}[\bar{f}%
(\varepsilon ^{\prime 2})]^{2}}}\mathrm{Re}\left[ \frac{1}{\sqrt{i\bar{f}%
\left( \varepsilon ^{\prime }\right) +\bar{\mu}}}\right] .
\end{gather*}%
The numerical solution of these equation gives consistently higher
transition temperatures than for the charge orders. Actually, the $T_{c}$
obtained correspond roughly to a twice larger coupling for the SC channel
(see also Eq. (\ref{sf_tdiag})). Qualitatively this can be explained by the
fact that for a certain wave vector, only two of four hot spots in a region
have nesting (and the other two are nested by the reversed wave vector),
while the superconducting pairing is the same in all four hot spots. This
leads to the conclusion that including Coulomb interaction is crucial for
obtaining $T_{CDW}>T_{c}$. As soon as the superconducting transition is
completely suppressed, one can consider the charge ordering independently,
as has been done in our present study.

\section{\label{Concl} Conclusions and comparison with experiments.}

Considering the spin-fermion model we have shown that contributions coming
from the regions away from the Fermi surface in the antinodal regions can
have an important influence on the formation of the charge order provided
the dispersion is sufficiently shallow (which is motivated by the existing
ARPES data\cite{Bi2201ARPES-1,Bi2201ARPES-2,Bi2212ARPES-1}). The leading
instability has been shown to be a d-wave Fermi surface distortion followed
at a lower temperature by a transition into a state with a d-form factor CDW
modulated with a vector directed along one of the BZ axes.

We have found that an overlap between the hot spots in the spin-fermion
model leads to strongly damped quasiparticles in the normal state in accord
with ARPES experiments. The transition temperatures $T_{Pom}$
and $T_{CDW}$ can be not far away from each other provided one takes into
account effects of the nesting emerging on the deformed Fermi surface.

This leads to the following qualitative picture of the charge order
formation:

$\bullet $ At $T_{Pom}\gtrsim T^{\ast }$ a $C_{4}$-symmetry breaking Pomeranchuk
transition occurs. It manifests itself in d-form factor deformation of the
Fermi surface (see Fig. \ref{figpom}) and a redistribution of the charge
between the oxygen orbitals inside the unit cell. This can lead to formation
of domains with different signs of the order parameter and different
orientations of the deformed Fermi surface thus concealing the C$_{4}$%
-symmetry breaking for bulk probes. 

$\bullet$ At $T_{CDW}<T_{Pom}$ the d-form factor CDW forms. The wave vector
is directed along one of the BZ axes depending on the sign of the
Pomeranchuk order parameter. The
absolute value of the CDW wave vector $Q_{CDW}$ decreases with hole doping. It generally exceeds $Q_{AN}$ and should be determined self-consistently by the interaction and parameters of the Fermi surface. As a result, no universal
relation between $Q_{CDW}$ and $Q_{AN}$ can be obtained.

$\bullet $ At $T_{CDW}$ the deformation of the Fermi surface should be
sufficiently large in order to deform the Fermi surface to a shape with
parts close to nesting like those in Fig. \ref{figpom} orthogonal to the
initial one. This type of the deformation can lead to quite high transition
temperatures of the order of $T_{Pom}$.

The picture arising from the spin fermion model is quite general without a
need of fine-tuning the parameters of the model or introducing additional
components/orders.

Our findings help to explain the results of recent experiments. The
Pomeranchuk deformation as a leading instability explains well the C$_{4}$
symmetry breaking at commensurate peaks in Fourier transformed STM data\cite%
{Bi2212STM-0}. Formation of domains with different types of the C$_{4}$%
-symmetry breaking is seen in STM experiments \cite{Bi2212STM-3} and can
also help explaining results of the transport measurements in YBCO\cite%
{y123Transp-2}. The results of the measurements of Ref. %
\onlinecite{y123Transp-2} show that the orientational transition preempts
formation of CDW only at low dopings. However, one can imagine that,
provided there are domains with different signs of the Pomeranchuk order,
this transition can be seen in bulk probes only if one of the orientations
is strongly preferred. It is thus possible that the C$_{4}$-symmetry
breaking at higher doping is not seen in the transport measurements due to a
rather small difference between the densities of the domains with the two
different orientations. This may also resolve the apparent contradiction to
the ARPES data \cite{Bi2201ARPES-1,Bi2212ARPES-1} always showing a C$_{4}$%
-symmetric Fermi surface.

The most important aspect of the Pomeranchuk order is that it explains the
robustness of the axial d-form factor CDW in the cuprates. We also note that
the organization of the CDW phase in the unidirectional domains is indeed
seen in STM\cite{Bi2212STM-3} and XRD\cite{y123REXS-4,HgXRD} measurements.
The coexistence of the CDW and Pomeranchuk order also allows one to resolve
a seeming contradiction to results obtained in experiments on quantum
oscillations \cite{seb1,seb2,sachQO}. Although the unidirectional CDW leads to an
open Fermi surface that does not support quantum oscillations, it has been
shown in Ref. \onlinecite{kivelson2011} that the simultaneous presence of a C%
$_{4}$-symmetry breaking can indeed close the Fermi surface leading to
quantum oscillations in high magnetic fields.

\begin{acknowledgments}
The authors gratefully acknowledges the financial support of the Ministry of
Education and Science of the Russian Federation in the framework of Increase
Competitiveness Program of NUST~\textquotedblleft MISiS\textquotedblright\
(Nr.~K2-2014-015).
\end{acknowledgments}

\appendix

\section{\label{app1a}Hole density on oxygen sites in ZRS picture}

Following Ref. \onlinecite{ZRS} we assume that, in the limit of a weak
tunneling between Cu and O atoms, doped holes entering the CuO$_{2}$ plane
occupy mainly O sites forming a bound singlet state with a hole sitting on a
Cu atom. In terms of the single band model derived in Ref. \onlinecite{ZRS},
this means that the double hole occupancy of a site should be interpreted as
presence of such a bound state. The wave function of this state centered
around a $Cu$ hole at site $i$ with coordinate $\mathbf{R}_{i}$ is given by
\begin{eqnarray}
&&|ZRS\rangle _{i}=\frac{|d_{i},\uparrow \rangle |\phi _{i},\downarrow
\rangle -|d_{i},\downarrow \rangle |\phi _{i},\uparrow \rangle }{\sqrt{2}},
\notag \\
&&|\phi _{i},\sigma \rangle =\sum_{\alpha }a_{\alpha }^{i}|O_{\alpha
},\sigma \rangle   \label{app1_zrs} \\
&\approx &\frac{1}{2}\left( |O_{x-}^{i},\sigma \rangle -|O_{x+}^{i},\sigma
\rangle +|O_{y-}^{i},\sigma \rangle -|O_{y+}^{i},\sigma \rangle \right) ,
\notag
\end{eqnarray}%
where the index $\alpha $ enumerates all oxygen sites in the $CuO_{2}$ plane
(we will use Greek indices for oxygen sites in what follows), $|d_{i}, \sigma \rangle$ denotes a state with a single hole at $Cu$ site with spin $\sigma$, $O_{x-(+)}$ stands for
the left (right) neighboring oxygen orbital and $O_{y-(+)}$ is the
lower(upper) one. As discussed in Ref. \onlinecite{ZRS}, the approximate
expression above is not a proper state to construct an orthonormal basis
because the neighboring states are not orthogonal to each other. We will
employ it merely for estimating the magnitude of coefficients $a_{\alpha
}^{i}$ in the final expressions, while taking into account in the derivation
only general properties of the $\phi _{i}$ states.

To calculate the physical properties of the holes on oxygen sites we have to
consider the action of the corresponding operators on ZRS states. We start
with writing an operator destroying a hole on $O$ site $\alpha $ with spin $%
\sigma $:

\begin{equation}
\hat{p}_{\alpha ,\sigma }|ZRS\rangle _{i}=\sum_{i}a_{\alpha
}^{i}|d_{i},-\sigma \rangle =\sum_{i}a_{\alpha }^{i}\hat{c}_{i,\sigma
}^{\dagger }|0\rangle ,  \label{b15}
\end{equation}%
where $\hat{c}_{i,\sigma }^{\dagger }$ is the creation operator of an
electron on the site $i$ with spin $\sigma $.

In the single-band model, a ZRS on the site $i$ corresponds to an unoccupied
site and the operator $\hat{p}_{\alpha ,\sigma }$ effectively creates
electrons in the single-band model on vacant sites. If the site was
occupied, however, the action of $\hat{p}_{\alpha ,\sigma }$ operator should
be identically zero to avoid the double occupancy. Therefore, one should
take into account this restriction by projecting out states with site $i$
occupied. These arguments lead us to the following operator form of $%
p_{\alpha ,\sigma }$ in the single-band representation:

\begin{gather}
\hat{p}_{\alpha ,\sigma }\equiv \frac{1}{\sqrt{2}}\sum_{i}a_{\alpha }^{i}%
\hat{c}_{i,\sigma }^{\dagger }\hat{\Pi}_{i}  \notag \\
\hat{\Pi}_{i}=(1-\hat{n}_{i,\uparrow })(1-\hat{n}_{i,\downarrow }),
\label{b16}
\end{gather}%
where $\hat{n}_{i,\sigma }=\hat{c}_{i,\sigma }^{\dagger }\hat{c}_{i,\sigma }$

In order to exclude the double occupancy of an arbitrary site we use the
Gutzwiller projection operator
\begin{eqnarray}
\hat{P}_{GW} &=&\prod_{i}\hat{P}_{GW}^{i},  \label{b17} \\
\hat{P}_{GW}^{i} &=&(1-\hat{n}_{i,\uparrow }\hat{n}_{i,\downarrow }).  \notag
\end{eqnarray}

The operator $\hat{P}_{GW}$ projects out states with the double electron
occupancy, such that in the remaining wave function unoccupied states can
occur only due to doping. To keep the wave function normalized we need to
divide all the averages obtained by a factor $\langle \hat{P}_{GW}\rangle $.
In what follows we will denote a normalized average of an operator $\hat{A}$
over a Gutzwiller projected state by $\langle \langle \hat{A}\rangle \rangle
$. Now we are in position to evaluate some averages that will be useful
later on

\begin{gather}
\langle \langle \hat{n}_{i,\sigma }\rangle \rangle =\frac{\langle \hat{P}%
_{GW}c_{i,\sigma }^{\dagger }c_{i,\sigma }\hat{P}_{GW}\rangle }{\langle \hat{%
P}_{GW}\rangle }=\frac{1-p}{2},  \notag \\
\hat{\Pi}_{i}=1-\hat{n}_{i,\uparrow }-\hat{n}_{i,\downarrow }+\hat{n}%
_{i,\uparrow }\hat{n}_{i,\downarrow }=2-\hat{n}_{i,\uparrow }-\hat{n}%
_{i,\downarrow }-\hat{P}_{GW}^{i},  \notag \\
\langle \langle \hat{\Pi}_{i}\rangle \rangle =\frac{\langle \hat{P}_{GW}\Pi
_{i}\hat{P}_{GW}\rangle }{\langle \hat{P}_{GW}\rangle }=p,  \label{b18}
\end{gather}%
where $p$ is the relative hole doping. We have used the fact that the normal
state is an eigenstate of the total particle number operator as well as the
uniformity of the normal state.

Now let us calculate the hole density on an oxygen site in the normal state:

\begin{eqnarray}
&&n_{\alpha ,\sigma }^{O}=\langle \langle \hat{p}_{\alpha ,\sigma }^{\dagger
}\hat{p}_{\alpha ,\sigma }\rangle \rangle =\frac{1}{2}\sum_{ij}(a_{\alpha
}^{i})^{\ast }a_{\alpha }^{j}\langle \langle \hat{\Pi}_{i}\hat{c}_{i,\sigma }%
\hat{c}_{j,\sigma }^{\dagger }\hat{\Pi}_{j}\rangle \rangle  \notag \\
&=&\frac{1}{2}\sum_{i}|a_{\alpha }^{i}|^{2}\langle \langle \hat{\Pi}_{i}(1-%
\hat{n}_{i,\sigma })\hat{\Pi}_{i}\rangle \rangle  \notag \\
&&+\frac{1}{2}\sum_{i\neq j}(a_{\alpha }^{i})^{\ast }a_{\alpha }^{j}\langle
\langle \hat{\Pi}_{i}\hat{c}_{i,\sigma }\hat{c}_{j,\sigma }^{\dagger }\hat{%
\Pi}_{j}\rangle \rangle  \notag \\
&=&\frac{1}{2}\sum_{i}|a_{\alpha }^{i}|^{2}\langle \langle \hat{\Pi}%
_{i}\rangle \rangle +\frac{1}{2}\sum_{i\neq j}(a_{\alpha }^{i})^{\ast
}a_{\alpha }^{j}\langle \langle \hat{\Pi}_{i}\hat{c}_{i,\sigma }\hat{c}%
_{j,\sigma }^{\dagger }\hat{\Pi}_{j}\rangle \rangle  \notag \\
&=&\frac{p}{2}\sum_{i}|a_{\alpha }^{i}|^{2}+\frac{1}{2}\sum_{i\neq
j}(a_{\alpha }^{i})^{\ast }a_{\alpha }^{j}\langle \langle \hat{\Pi}_{i}\hat{c%
}_{i,\sigma }\hat{c}_{j,\sigma }^{\dagger }\hat{\Pi}_{j}\rangle \rangle .
\label{b18b}
\end{eqnarray}

As the system is assumed to be uniform, the mean hole density should not
depend on $\alpha $. Taking into account the orthonormality conditions $%
\sum_{\alpha }(a_{\alpha }^{i})^{\ast }a_{\alpha }^{j}=\delta _{ij}$ one has
\begin{equation}
n_{\sigma }^{O}=\frac{1}{2N}\sum_{\alpha }n_{\alpha ,\sigma }^{O}=\frac{p}{2}%
\cdot \frac{1}{2N}\sum_{i}1=\frac{p}{4}.  \label{b19}
\end{equation}%
Now let us consider the case when the bond order in the single band model is
present. In this case, the density of the oxygen holes changes due to the
second term in Eq. (\ref{b18b}). The unprojected mean-field state is
characterized by a change of the expectation value of the bond operator
$\delta \langle \hat{c}_{i,\sigma }\hat{c}_{j,\sigma }^{\dagger }\rangle \equiv \langle \hat{c}_{i,\sigma }\hat{c}_{j,\sigma }^{\dagger }\rangle _{CO}$ in the ordered phase. To calculate corresponding change in the projected
average $\delta \langle \langle \hat{\Pi}_{i}\hat{c}_{i,\sigma }\hat{c}%
_{j,\sigma }^{\dagger }\hat{\Pi}_{j}\rangle \rangle$ we note that the
operators $\hat{c}_{i,\sigma }\hat{c}_{j,\sigma }^{\dagger }$ transfer an
electron from site $i$ to site $j$. As the double-occupied sites are
projected out, the site $j$ should be empty, while the site $i$ is occupied
by an electron. Assuming that these two conditions are uncorrelated we can
reduce the influence of the projection to multiplication by projection
factor $p(1-p)\approx p$ for small doping. Thus, one obtains
\begin{equation}
\delta n_{\alpha ,\sigma }^{O}\approx p\frac{1}{2}\sum_{i\neq j}(a_{\alpha
}^{i})^{\ast }a_{\alpha }^{j}\langle \hat{c}_{i,\sigma }\hat{c}_{j,\sigma
}^{\dagger }\rangle _{CO},  \label{b20}
\end{equation}%
where $\langle ... \rangle_{CO}$ is the contribution due to charge ordering in the corresponding average. Now we rewrite $\langle \hat{c}_{i,\sigma }\hat{c}_{j,\sigma }^{\dagger
}\rangle _{CO}$ in the momentum space
\begin{equation}
\langle \hat{c}_{i,\sigma }\hat{c}_{j,\sigma }^{\dagger }\rangle _{CO}=-e^{-i%
\mathbf{Q(R}_{i}\mathbf{+R}_{j}\mathbf{)}/2}\sum_{\mathbf{k}}W_{\mathbf{Q}%
}(k)e^{i\mathbf{Q}(\mathbf{R}_{i}-\mathbf{R}_{j})}.  \label{b21}
\end{equation}%
For the pure d-form factor $\cos k_{x}-\cos k_{y}$ only the
nearest-neighbor correlation functions do not vanish. Moreover, the
contribution of non-nearest neighbor correlators to the hole density on the
oxygen sites is suppressed because $|a_{\alpha }^{i}|$ decreases rapidly for
sites not adjacent to the oxygen site $\alpha $ (\textit{e.g.} $0.08$ for
next nearest neighboring $Cu$ sites against $0.48$ for the nearest ones).
This means that we can use the approximate expression for the ZRS
wave function (\ref{app1_zrs}) to finally obtain
\begin{equation}
n_{\alpha ,\sigma }^{O}\approx \frac{p}{4}+\frac{p}{8}\langle c_{i+1,\sigma
}^{\dagger }c_{i,\sigma }+c_{i,\sigma }^{\dagger }c_{i+1,\sigma }\rangle
_{CO},  \label{b22}
\end{equation}%
where $i$ and $i+1$ are the Cu sites nearest to the oxygen site $\alpha $.

\section{\label{app0}Hubbard-Stratonovich transformation.}

Using the vectors $\Psi $ and $\Psi ^{\dagger }$ introduced in Eq. (\ref%
{ap4b}) is very convenient when the energies of the charge orders and
superconductivity are close to each other. In addition to the hermitian
conjugate $\Psi ^{\dagger }$ of the vector $\Psi $ we introduce a
\textquotedblleft charge\textquotedblright\ conjugate field~$\bar{\Psi}$
defined as
\begin{equation}
\bar{\Psi}=\left( C\Psi \right) ^{t}\quad \text{\textnormal{with}}\quad
C=\left(
\begin{array}{cc}
0 & i\sigma _{2} \\
-i\sigma _{2} & 0%
\end{array}%
\right) _{\tau }=-\tau _{2}\sigma _{2}.  \label{a9a}
\end{equation}%
The matrix $C$ satisfies the relations $C^{t}C=1$ and $C=C^{t}$.

It is clear that
\begin{equation}
\bar{\Psi}=\Psi ^{\dagger }\tau _{3}\ ,  \label{a9b}
\end{equation}%
where
\begin{equation*}
\tau _{3}=\left(
\begin{array}{cc}
1 & 0 \\
0 & -1%
\end{array}%
\right)
\end{equation*}

The notion of the charge conjugation is naturally extended to arbitrary
matrices $M\left( X,X^{\prime }\right) $ as
\begin{equation}
\left( \bar{\Psi}\left( X\right) M\left( X,X^{\prime }\right) \Psi \left(
X^{\prime }\right) \right) =-\left( \bar{\Psi}\left( X^{\prime }\right) \bar{%
M}\left( X^{\prime },X\right) \Psi \left( X\right) \right) .  \label{ap5}
\end{equation}%
It is easy to see that
\begin{equation}
\bar{M}\left( X,X^{\prime }\right) =CM^{t}\left( X^{\prime },X\right)
C^{t}\equiv CM^{T}\left( X,X^{\prime }\right) C^{t}\ .  \label{ap6}
\end{equation}%
Matrices satisfying the relation
\begin{equation}
\bar{M}=-M  \label{a12a}
\end{equation}%
are \emph{anti-selfconjugated}.

Using the field~$\Psi $ we rewrite Eq. (\ref{b2}) in the form
\begin{equation}
S_{\mathrm{0}}\left[ \Psi \right] =\int \bar{\Psi}\left( X\right) \mathcal{H}%
_{0}\Psi \left( X\right) dX,  \label{ap9}
\end{equation}%
where the operator $\mathcal{H}_{0}=I_{\sigma }\otimes H_{0},$ $I_{\sigma }$
is the unit matrix in the spin blocks, and $H_{0}$ is determined by Eq. (\ref%
{k3}).

The term $S_{\mathrm{\psi }}$, Eq. (\ref{b2a}), takes the form
\begin{equation}
S_{\mathrm{\psi }}=\lambda \int \bar{\Psi}\left( X\right) \vec{\sigma}^{t}%
\vec{\phi}\left( X\right) \Psi \left( X\right) dX,  \label{d1}
\end{equation}%
whereas $S_{\mathrm{int}}$, Eq. (\ref{b4}), can be written as
\begin{eqnarray}
&&S_{\mathrm{int}}\left[ \Psi \right] =-\frac{\lambda ^{2}}{2}\int D_{%
\mathrm{0}}\left( X-X^{\prime }\right)   \label{b8} \\
&&\times \left( \bar{\Psi}\left( X\right) \vec{\sigma}^{t}\Psi \left(
X\right) \right) \left( \bar{\Psi}\left( X^{\prime }\right) \vec{\sigma}%
^{t}\Psi \left( X^{\prime }\right) \right) dXdX^{\prime }  \notag
\end{eqnarray}%
The Coulomb interaction $S_{\mathrm{c}}$, Eq. (\ref{b2b}), reads now
\begin{eqnarray}
&&S_{\mathrm{c}}=\frac{1}{2}\int V_{\mathrm{c}}\left( X-X^{\prime }\right)
\label{d3} \\
&&\times \left( \bar{\Psi}\left( X\right) \tau _{3}\Psi \left( X\right)
\right) \left( \bar{\Psi}\left( X^{\prime }\right) \tau _{3}\Psi \left(
X^{\prime }\right) \right) dXdX^{\prime }  \notag
\end{eqnarray}%
and the partition function $Z$ is given as before by Eq. (\ref{b6}).

Eqs. (\ref{b5}, \ref{b6}, \ref{d1}, \ref{ap9}, \ref{b8}, \ref{b9}) fully
define the model under study.

As usual\cite{Efetov2013}, in order to reduce the integration over the
fermionic fields to integration over slowly varying in space and time order
parameters one has to single out slowly varying pairs of the fermionic
fields. In the model under consideration, one can expect singlet
superconductivity and charge orders, whereas a triplet superconductivity and
spin orders are less favorable energetically.

There are $2$ equivalent possibilities to form pairs in the term $S_{\mathrm{%
int}}$, Eq. (\ref{b8}) and we write the low-energy part of $S_{\mathrm{int}}%
\left[ \Psi \right] $ as%
\begin{eqnarray}
&&S_{\mathrm{int}}\left[ \Psi \right] \rightarrow \lambda ^{2}\mathrm{tr}%
\int D\left( X-X^{\prime }\right)  \label{b9} \\
&&\times \vec{\sigma}^{t}\left( \Psi \left( X\right) \bar{\Psi}\left(
X^{\prime }\right) \right) \vec{\sigma}^{t}\left( \Psi \left( X^{\prime
}\right) \bar{\Psi}\left( X\right) \right) dXdX^{\prime }.  \notag
\end{eqnarray}%
Pairing two fermionic fields $\Psi $ at equal variables $X$ would imply
existence of a spin density wave but it is assumed from the beginning that
the antiferromagnetic order is destroyed and one can check that additional
spin structures do not appear on the metallic side. Therefore, only pairs in
the brackets remain relevant in Eq. (\ref{b9}). They give, depending on
parameters of the model, singlet superconductivity or the charge order with
a certain modulation vector $\mathbf{Q}$. The value of $\mathbf{Q}$ is also
determined by the parameters of the Hamiltonian. Within the standard SF
model, the vector $\mathbf{Q}$ is given by the distance between the hotspots
\cite{MetSach,Efetov2013}. However, we will see that in the extended SF
model introduced in the present paper, a charge modulation with $\mathbf{Q}%
=0 $ can be more favorable in the limit of a \textquotedblleft
shallow\textquotedblright\ spectrum near the antinodes. Such a pairing leads
to a reconstruction of the Fermi surface.

The Coulomb interaction is also important for obtaining this kind of the
charge order. Writing relevant slow pairs in the Coulomb interaction $S_{%
\mathrm{c}}$, Eq. (\ref{d3}), one obtains two types of the contributions%
\begin{eqnarray}
&&S_{\mathrm{c}} \rightarrow \frac{1}{2}\int V_{\mathrm{c}}\left(
X-X^{\prime }\right) \left( \bar{\Psi}\left( X\right) \tau _{3}\Psi \left(
X\right) \right)  \notag \\
&&\times \left( \bar{\Psi}\left( X^{\prime }\right) \tau _{3}\Psi \left(
X^{\prime }\right) \right) dXdX^{\prime }  \notag \\
&&-\mathrm{tr}\int V_{\mathrm{c}}\left( X-X^{\prime }\right) \tau _{3}\left(
\Psi \left( X\right) \bar{\Psi}\left( X^{\prime }\right) \right)  \label{d4}
\\
&&\times \tau _{3}\left( \Psi \left( X^{\prime }\right) \bar{\Psi}\left(
X\right) \right) dXdX^{\prime }.  \notag
\end{eqnarray}%
The second\ term in Eq. (\ref{d4}) is analogous to the one in Eq. (\ref{b9}%
), while the first term stands for the classical part of the Coulomb
interaction.

As soon as the interaction terms are written in terms of products of slow
varying pairs of the fermionic fields, one can decouple the interaction
integrating over slowly varying matrix fields $\mathcal{M}\left( X,X^{\prime
}\right) $ having the same symmetry as the pairs $\Psi \left( X\right) \bar{%
\Psi}\left( X^{\prime }\right) \tau _{3}$ do. As the triplet superconducting
pairing and spin density waves are less favorable we perform the decoupling
using the matrices having numbers instead of spin blocks. After the
decoupling is performed, the effective Lagrangian is quadratic in the
fermionic fields and one can integrate out the latter.

As a result of all these standard manipulations \cite{Efetov2013}, one comes
to the following expression for the partition function $Z$%
\begin{eqnarray}
Z &=&\int \exp \Big[\frac{1}{2}\mathrm{Tr}\ln \left( \mathcal{H}_{0}-%
\mathcal{M}\right) \Big]  \label{b10} \\
&&\times \exp \Big[-\frac{1}{4}\mathrm{Tr}\left[ \mathcal{M}\left(
X,X^{\prime }\right) \hat{\Pi}_{\mathrm{s}}^{-1}\mathcal{M}\left( X^{\prime
},X\right) \right] \Big]D\mathcal{Q}.  \notag
\end{eqnarray}%
In Eq. (\ref{b10}) operator $\hat{\Pi}_{\mathrm{s}}$ acts on an arbitrary
matrix function $P\left( X,X^{\prime }\right) $ as%
\begin{eqnarray}
&&\hat{\Pi}_{\mathrm{s}}P\left( X,X^{\prime }\right) =3\lambda
^{2}D(X-X^{\prime })P\left( X,X^{\prime }\right)  \notag \\
&&-V_{\mathrm{c}}\left( X-X^{\prime }\right) \tau _{3}P\left( X,X^{\prime
}\right) \tau _{3}  \label{b11a} \\
&&+\frac{1}{2}\delta \left( X-X^{\prime }\right) \int V_{\mathrm{c}}\left(
X-X_{1}\right) \mathrm{trtr}_{\sigma }\left[ \tau _{3}P\left(
X_{1},X_{1}\right) \right] dX_{1},  \notag
\end{eqnarray}%
where $\mathrm{tr}$ is trace in the Gor'kov-Nambu space and $\mathrm{tr}%
_{\sigma }$ is trace in the spin space.

Mean field equations are simply saddle point equations for the integral over
$\mathcal{M}$ in Eq. (\ref{b10}). Minimizing the exponent in Eq. (\ref{b10})
we come to Eqs. (\ref{k1}-\ref{k3}) with the order parameter $M$ given by
Eq. (\ref{ap4da}-\ref{k3}).

\section{\label{app2}Existence of $\protect\mu_{cr}$ for SF model}

We consider a linearized equation for the CDW order parameter $\bar{W}(\bar{%
\varepsilon}_{n})$. Assuming the CDW wave vector to be small we can
decompose the R.H.S. and see at which $\bar{\mu}$ the modulation with $Q=0$
becomes favorable. We write the equation in the form
\begin{eqnarray}
&&\bar{W}(\bar{\varepsilon}_{n})=\frac{i\cdot 0.75\bar{T}}{2}\sum_{\varepsilon
_{n}^{\prime }}\frac{\bar{W}(\bar{\varepsilon}_{n})}{\sqrt{\bar{\Omega}(\bar{%
\varepsilon}_{n}-\bar{\varepsilon _{n}^{\prime }})+\bar{a}}}  \notag \\
&&\times \left[ \frac{\mathrm{sgn}(\mathrm{Re}[f(\bar{\varepsilon}%
_{n}^{\prime })])}{\left( i\bar{f}(\bar{\varepsilon}_{n}^{\prime })+\bar{\mu}%
\right) ^{3/2}}+\frac{Q^{2}}{4}\frac{\mathrm{sgn}(\mathrm{Re}[f(\bar{%
\varepsilon}_{n}^{\prime })])}{\left( i\bar{f}(\bar{\varepsilon}_{n}^{\prime
})+\bar{\mu}\right) ^{5/2}}\right] .  \label{b12}
\end{eqnarray}

In our numerical results the imaginary part of the solution has always been
much smaller than the real one and concentrated at low frequencies only. For
purely real order parameter one has
\begin{gather}
\bar{W}(\bar{\varepsilon}_{n})=\frac{-0.75\bar{T}}{2}\sum_{\varepsilon
_{n}^{\prime }}\frac{\bar{W}(\bar{\varepsilon}_{n})\mathrm{sgn}(\mathrm{Re}[f(\bar{%
\varepsilon}_{n}^{\prime })])}{\sqrt{\bar{\Omega}(\bar{\varepsilon}_{n}-\bar{%
\varepsilon _{n}^{\prime }})+\bar{a}}}  \notag \\
\times \mathrm{Im}\left[ \frac{1}{\left( i\bar{f}(\bar{\varepsilon}%
_{n}^{\prime })+\bar{\mu}\right) ^{3/2}}+\frac{Q^{2}}{4}\frac{1}{\left( i%
\bar{f}(\bar{\varepsilon}_{n}^{\prime })+\bar{\mu}\right) ^{5/2}}\right] .
\label{b13}
\end{gather}

Then, one sees that having $Q\neq 0$ is favorable provided the imaginary
parts of the two terms in the R.H.S. of Eq. (\ref{b13}) have the same signs.
For the first term, $\left( i\bar{f}(\bar{\varepsilon}_{n}^{\prime })+\bar{%
\mu}\right) ^{-3/2}$, the sign of the imaginary part is always $-\mathrm{sgn}%
(\mathrm{Re}[f(\bar{\varepsilon}_{n}^{\prime })])$ and therefore the R.H.S.
of Eq. (\ref{b13}) is always positive at $Q=0$. As concerns the contribution
coming from the second term, it can change the sign of the imaginary part.
For estimation we assume that for $Q=0$ to be favorable we need the signs of
imaginary parts of the two terms to be different for every frequency, even
for the lowest one. Then, one has
\begin{equation}
\mathrm{Im}\left[ \frac{1}{\left( i\bar{f}(\bar{\varepsilon}_{n}^{\prime })+%
\bar{\mu}_{cr}\right) ^{5/2}}\right] =0.  \label{b14}
\end{equation}%
Assuming $\mathrm{Im}[\bar{f}(\pi \bar{T})]$ be small we have the condition $%
\bar{f}(\pi \bar{T})/\bar{\mu}_{cr}=\tan (2\pi /5)\approx 3.08$. For $\bar{f}%
(\pi \bar{T})=\pi T$ one has $(\mu /T_{Pom})_{cr}\approx 1.02$ close to the
exact result, Eq. (\ref{k13a}) for the simplified model. However, taking
into account in Eq. (\ref{b14}) only the lowest frequency one underestimates
the value $\left( \mu /T_{Pom}\right) $. In SF model, substantially higher
values $\bar{f}(\pi \bar{T})$ can also be important leading eventually to
higher values of $(\mu /T_{Pom})_{cr}$. This is the reason why the numerical
results for this quantity obtained in Subsection \ref{PomSF} (see Fig. (\ref%
{fig5})) are considerably higher.


\begin{thebibliography}{99}
\bibitem{fexpand} Actually, expanding the function $\bar{I}\left( q\right) $
in $q$ one obtains $q^{2}\int_{-\infty }^{\infty }\left[ -8\frac{\sinh
(x^{2}-\kappa )}{\cosh ^{3}(x^{2}-\kappa )}-\frac{16}{3}x^{2}\frac{1-2\sinh
^{2}(x^{2}-\kappa )}{\cosh ^{4}(x^{2}-\kappa )}\right] dx$. However, the
integrating the second term by parts one comes immediately to Eq. (\ref{k13}%
).
\end{thebibliography}
\end{document}